\documentclass[preprint,showpacs,citeautoscript,secnumarabic,longbibliography,nobibnotes, amsmath,amssymb,aps,prl,floatfix, superscriptaddress
]{revtex4-2}

\usepackage{setspace}
\usepackage{graphicx}
\usepackage{dcolumn}
\usepackage{bm}
\usepackage{hyperref}

\usepackage{multirow}
\usepackage{soul}
\usepackage{color}
\usepackage{xspace}
\usepackage{siunitx}
\usepackage{amsmath}

\newcommand{\PbSnSe}{Pb$_{1-x}$Sn$_x$Se\xspace}
\newcommand{\PbSnTe}{Pb$_{1-x}$Sn$_x$Te\xspace}
\newcommand{\Xbar}{$\bar{X}$\xspace}

\newcommand{\GXG}{$\bar{\Gamma}-\bar{X}-\bar{\Gamma}$\xspace}

\newcommand{\SM}{Supplementary Information\xspace}

\newcommand{\KTH}{KTH Royal Institute of Technology, Applied Physics, Quantum Light and Matter Physics, AlbaNova universitetscentrum, 106 91 Stockholm, Sweden}
\newcommand{\Groningen}{Zernike Institute for Advanced Materials, University of Groningen, 9747 AG Groningen, The Netherlands}
\newcommand{\Chalmers}{Department of Physics and Astronomy, Chalmers University of Technology, 412 96 Gothenburg, Sweden}
\newcommand{\UU}{Department of Chemistry – \AA ngstr\"om laboratories, Uppsala University, 75120 Uppsala, Sweden}

\hypersetup{colorlinks=true, linkcolor=blue, citecolor=blue, urlcolor=blue, unicode=false}

\begin{document}

\preprint{}

\author{Magnus~H.~Berntsen}
\thanks{Corresponding authors: mhbe@kth.se, oscar@kth.se}
\affiliation{\KTH}
\author{Wanyu.~Chen}
\affiliation{\KTH}
\author{Dibya.~Phuyal}
\affiliation{\KTH}
\author{Jianyu~Wu}
\affiliation{\KTH}
\author{Antonija~Grubi\v{s}i\'{c}-\v{C}abo}
\affiliation{\KTH}
\affiliation{\Groningen}
\author{Maciej~Dendzik}
\affiliation{\KTH}
\author{Qinda~Guo}
\affiliation{\KTH}
\author{Andreas~Orthaber}
\affiliation{\UU}
\author{Cong~Li}
\affiliation{\KTH}
\author{Richard~Matthias~Geilhufe}
\affiliation{\Chalmers}
\author{Jonas~Weissenrieder}
\affiliation{\KTH}
\author{Oscar~Tjernberg}
\thanks{Corresponding authors: mhbe@kth.se, oscar@kth.se}
\affiliation{\KTH}

\title{Optically driven ultrafast lattice contraction in a topological crystalline insulator}



\begin{abstract}
The ability to dynamically control topological order remains a central challenge in the exploration of quantum materials. Here, we investigate the lattice and electronic response of the topological crystalline insulator (TCI) \PbSnSe to ultrafast optical excitation using time- and angle-resolved photoemission spectroscopy (tr-ARPES) in combination with X-ray diffraction (XRD) and first-principles calculations. Static XRD measurements establish the lattice constant as the key control parameter for the topological transition, with a critical value $a_\mathrm{c}=6.073$~\AA\ separating the normal-insulator and topological phases. Despite generating electronic temperatures far above the topological-to-normal transition temperature $T_\mathrm{c}$, optical excitation in the TCI phase does not induce a phase transition. Instead, it drives an ultrafast lattice contraction of $\Delta a \approx -0.45$~pm within less than 1~ps that counteracts thermal expansion and prevents the system from reaching $a_\mathrm{c}$. The optical excitation therefore pushes the system deeper into the topological phase, providing a direct explanation for the absence of an optically driven transition. Total-energy calculations reveal that this behavior originates from the band inversion in the topological regime, which reverses the character of the photoinduced charge transfer and transiently strengthens atomic bonding. These findings identify band inversion as a mechanism that fundamentally alters the nonequilibrium lattice response, with general implications for ultrafast optical control of topological materials.

\end{abstract}

\maketitle


\section{Introduction}
The discovery of topological phases of matter has reshaped our understanding of electronic materials, demonstrating that global properties of band structures can give rise to robust, symmetry-protected boundary states \cite{Laughlin1981,Halperin1982,Thouless1982,Haldane1988,Hatsugai1993,Kane2005b,Schnyder2008,Kitaev2009}. Among these systems, topological crystalline insulators (TCIs)~\cite{Fu2011} constitute a particularly intriguing class, in which crystalline symmetries -- such as mirror, rotation, or glide symmetries -- protect metallic surface states. In contrast to time-reversal-symmetric topological insulators (TIs) \cite{Bernevig2006,Konig2007,Fu2007a,Moore2007,Hsieh2008,Roy2009}, the topological properties of TCIs depend explicitly on the underlying lattice symmetry, offering opportunities to manipulate the topological phase by selectively breaking these symmetries.

Ultrafast optical excitation with femtosecond laser pulses provides a powerful route to dynamically modify the lattice of solids. Through the excitation of coherent phonons, acoustic strain waves, or electronically driven structural transitions, the atomic structure can be transiently distorted on ultrafast timescales, leading to corresponding changes in electronic, optical, and magnetic properties \cite{Thomsen1986,Zeiger1992,Cavalleri2001,Rini2007,Mankowsky2014,Afanasiev2021}. In the context of TCIs, such optically induced lattice perturbations are particularly compelling, as they offer a pathway toward active and potentially reversible control of topological order. More fundamentally, the inverted electronic structure characteristic of topological phases may itself influence the nonequilibrium lattice response, enabling photoexcitation to induce unconventional atomic bond dynamics.

\begin{figure}
\includegraphics[width=0.6\linewidth]{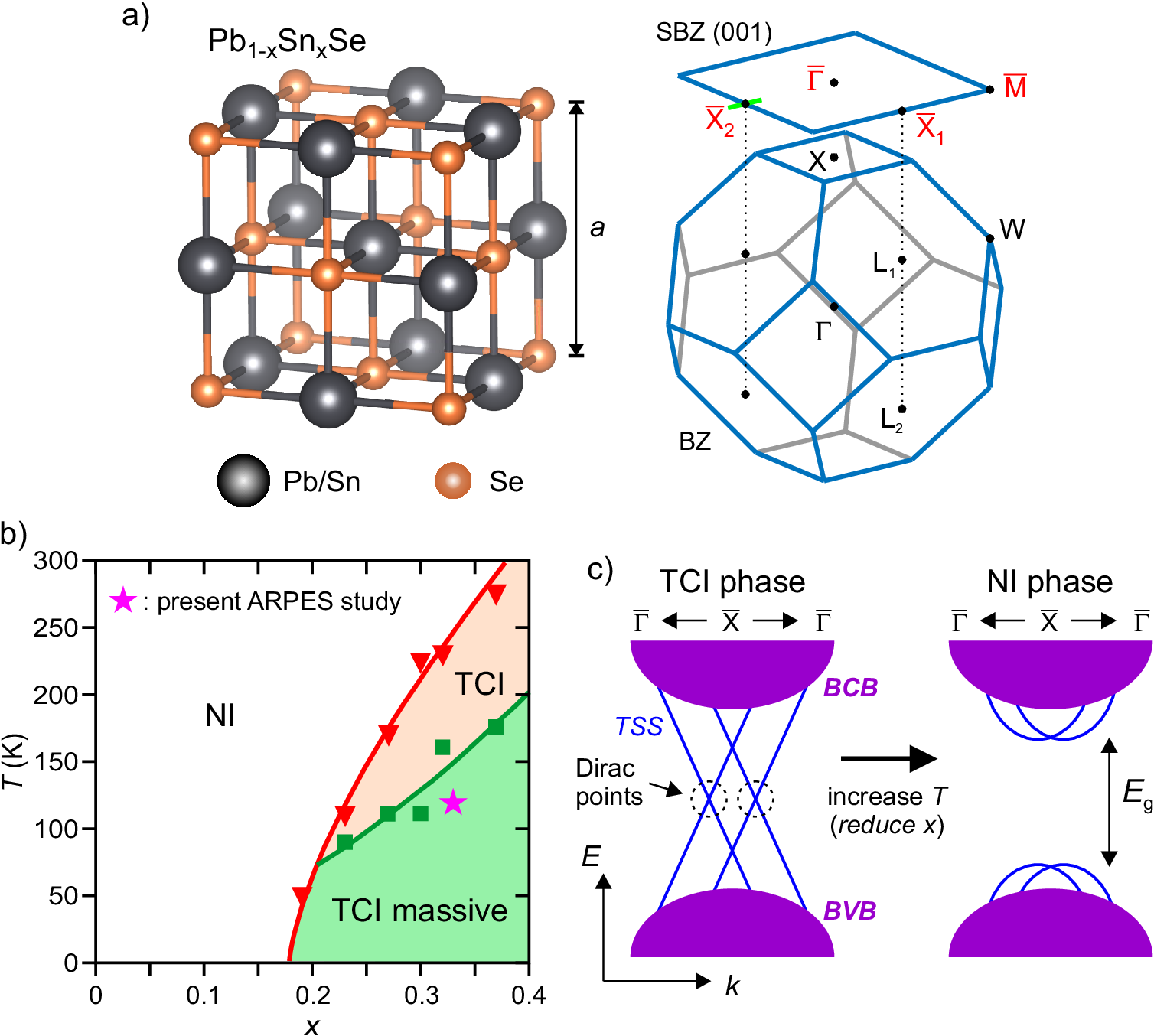}
\caption{\textbf{Evolution of the topological crystalline insulator phase in \PbSnSe.} a) Rock-salt crystal structure of \PbSnSe and schematic drawing of bulk Brillouin zone (BZ) as well as the projected surface Brillouin zone (SBZ) on the (001) crystal face. b) Topological phase diagram of \PbSnSe as a function of Sn content ($x$) and sample temperature ($T$) (adapted from ref.~\onlinecite{Wojek2015}). c) Schematic drawing of the energy--momentum dispersion of the topological surface state (TSS) in the TCI and NI phases, respectively. In the TCI phase, the TSS connects the bulk valence band (BVB) to the bulk conduction band (BCB). Transitioning from the TCI to the NI phase, either by a decrease in $x$ or an increase in $T$, opens an energy gap $E_\mathrm{g}$ in the TSS at the Dirac points.}
\label{fig1A}
\end{figure}

A prototypical TCI is the narrow-bandgap semiconductor \PbSnSe~\cite{Dziawa2012,Polley2014}, which adopts a cubic crystal structure for Sn concentrations ($x$) below 40\% and temperatures ($T$) ranging from room temperature to cryogenic values~\cite{Preier1979,Szczerbakow1994}. Figure~\ref{fig1A}a shows the crystal structure and the corresponding bulk Brillouin zone (BZ), as well as the surface Brillouin zone (SBZ) associated with the (001) termination of \PbSnSe. The material undergoes a topological phase transition between a normal insulator (NI) with a positive band gap and a TCI with an inverted band gap as a function of $x$ and $T$~\cite{Wojek2014}. The signature of this transition is the appearance or disappearance of Dirac-like topological surface states (TSS) within the bulk band gap, as illustrated in Fig.~\ref{fig1A}c. The corresponding $x$--$T$ phase diagram (Fig.~\ref{fig1A}b) has been extensively mapped by angle-resolved photoemission spectroscopy (ARPES) under static conditions~\cite{Wojek2014,Wojek2015}. These studies demonstrate that, in addition to symmetry breaking, the system can be tuned between the normal and topological phases by varying either the chemical composition or the temperature.

From an ultrafast perspective, however, the chemical composition is fixed, whereas the temperature can be modified on short timescales through laser-induced lattice heating. Following optical excitation, the electronic system relaxes via electron–phonon coupling, leading to a transient increase in lattice temperature. Despite this, optical excitation of Pb$_{0.79}$Sn$_{0.21}$Se \cite{Mogi2024} shows no evidence of an electronically or thermally driven TCI--NI phase transition. By contrast, Bi doping ($\mathrm{Bi}_y(\mathrm{Pb}_{1-x}\mathrm{Sn}_x)_{1-y}\mathrm{Se}$) introduces a rhombohedral lattice distortion that breaks inversion symmetry and converts the system into a strong time-reversal-invariant ($Z_2$) TI rather than a TCI~\cite{Mandal2017,Mogi2024}. In this case, an optically induced uniaxial lattice expansion along the out-of-plane direction has been shown to drive a $Z_2$ TI–to–NI phase transition~\cite{Mogi2024}.

For \PbSnSe in the TCI phase, the apparent robustness of the topological phase against optical excitation raises a key question: why does optical excitation fail to drive the system across the phase boundary despite the substantial energy deposited into the electronic and lattice subsystems? Addressing this question requires elucidating the microscopic origin of the TCI phase diagram, as well as determining the detailed electronic and lattice responses to optical excitation. While both $x$ and $T$ influence the NI--TCI transition, their effects are fundamentally different: alloying through $x$ modifies the spin--orbit coupling (Sn substitutes Pb) and bond lengths (Vegard's law), whereas temperature primarily affects lattice dynamics and thermal expansion or contraction. Whether, and under what conditions, an optically induced TCI–to–NI transition can be realized in \PbSnSe therefore remains an open question.

Here, we address these issues through systematic investigations of the $x$--$T$ dependence of the lattice constant in \PbSnSe, together with measurements of the response of the topological surface state to ultrafast optical excitation. Our results establish the lattice constant as the key tuning parameter for the NI–to–TCI transition, with crystal-field splitting providing the driving mechanism for band inversion in this material system. A well-defined critical lattice constant, $a_\mathrm{c}$, marks the topological phase boundary. Supported by band-structure and total-energy calculations, we show that optical excitation of \PbSnSe in the TCI phase induces an electronically driven lattice contraction that counteracts thermal expansion and prevents the system from reaching $a_\mathrm{c}$. The calculations further reveal that this behavior originates from the band inversion itself, which drives the system away from the phase boundary and deeper into the TCI phase. More broadly, the optically induced lattice contraction enabled by band inversion may represent a general nonequilibrium response of band-inverted materials.

\section{Results}

\subsection*{Lattice constant controls the equilibrium topological phase}

\begin{figure}
    \centering
    \includegraphics[width=0.6\linewidth]{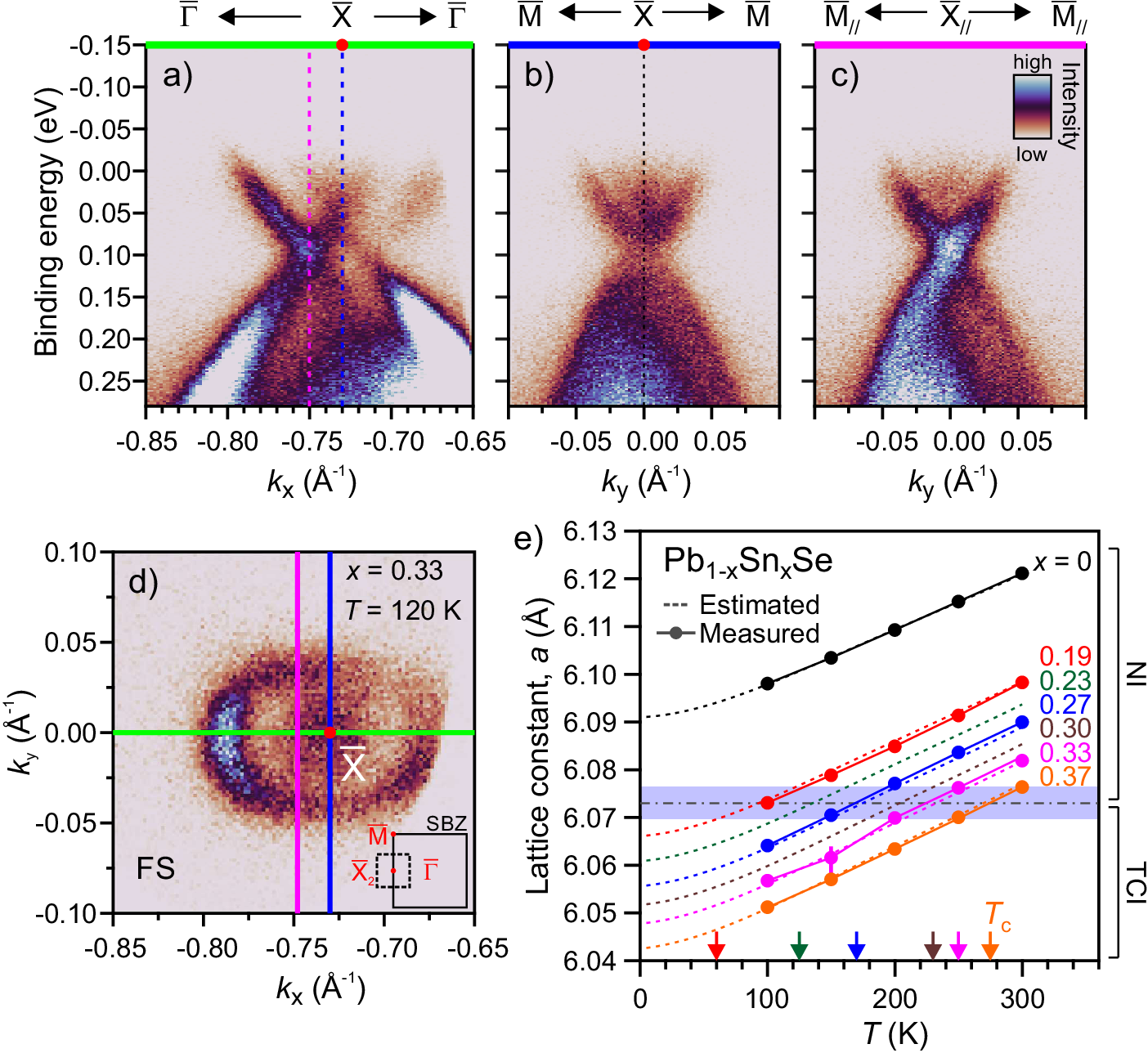}
    \caption{\textbf{Lattice constant determines topological phase.} a)--c) Energy-momentum dispersion of the TSS at \Xbar, measured at $T=120$~K by ARPES using 10.8 eV photons. The sample has $x=0.33$. The momentum positions of the cuts are given by the green, blue and purple lines in d). $\bar{\mathrm{M}}_{\parallel}$--$\bar{\Gamma}_{\parallel}$--$\bar{\mathrm{M}}_{\parallel}$ indicates a cut parallel to the $\bar{\mathrm{M}}$--$\bar{\Gamma}$--$\bar{\mathrm{M}}$ direction, intersecting the Dirac point of the TSS. d) Experimental constant energy contour of the TSS at the \Xbar point in the SBZ. e) Composition and temperature dependence of the lattice constant ($a$) in \PbSnSe. The different values of $x$ are given as labels next to the solid lines. Dashed lines are calculated values for $a$ using a two-dimensional empirical model function $a(x,T)$. Critical transition temperatures ($T_\mathrm{c}$) for the NI--TCI transition for the different compositions are marked on the $T$-axis with arrows in matching colors. The dashed horizontal line marks the critical value $a_\mathrm{c}=6.073$~\AA\xspace where the NI--TCI transition occurs. Vertical error bars are smaller than the data markers, except for the data point representing $x=0.33$ at $t=150$~K.}
    \label{fig1B}
\end{figure}

To elucidate the mechanism underlying the topological phase transition, we first systematically map the lattice constant of \PbSnSe over a broad range of compositions ($x$) and temperatures ($T$). The results (Fig.~\ref{fig1B}e) reveal an approximately linear temperature dependence of the lattice constant, with only a slight reduction in the slope ($\mathrm{d}a/\mathrm{d}T$) below $\sim 200$~K. The combined $x$–$T$ dependence is well captured by a model function $a(x,T)$ (dashed lines in Fig.~\ref{fig1B}e), constructed by combining the composition dependence of $a$ at room temperature with the previously reported temperature dependence for $x = 0$ \cite{Preier1979} (see Methods for details).

The equilibrium NI–TCI transition temperature $T_\mathrm{c}$ as a function of composition has previously been determined by temperature-dependent ARPES through the closing of the bulk band gap $E_\mathrm{g}$ and the emergence of gapless surface states near \Xbar in the (001) SBZ~\cite{Wojek2014}. As an example, Fig.~\ref{fig1B}a–d show static ARPES data (this work) for a sample with $x = 0.33$ measured in the TCI phase at $T = 120$~K ($T_\mathrm{c} = 250$~K), illustrating the presence of a gapless topological surface state. The critical temperatures for different compositions are indicated by color-coded arrows in Fig.~\ref{fig1B}e.

Remarkably, for all measured and estimated curves, the lattice constant at $T_\mathrm{c}$ falls within a narrow interval $\delta a = 0.68$~pm (shaded area in Fig.~\ref{fig1B}e) centered on a critical value $a_\mathrm{c} = 6.073$~\AA. This observation establishes a simple and direct criterion for the topological phase: $a > a_\mathrm{c}$ corresponds to the normal-insulator phase, whereas $a < a_\mathrm{c}$ corresponds to the TCI phase.

\subsection*{Photoexcitation leaves the topological surface state intact}

Our XRD results show that both $x$ and $T$ primarily influence the system through changes in the lattice constant, such that the static phase diagram in Fig.~\ref{fig1A}b can be parameterized in terms of $a$ alone (see right vertical axis in Fig.~\ref{fig1B}e). 
Since optical excitation ultimately heats the lattice, the near-surface region excited by an optical pump pulse is expected to undergo an isotropic lattice expansion, increasing $a$ and potentially driving the system from the TCI to the NI phase.

\begin{figure*}
\includegraphics[width=0.9\textwidth]{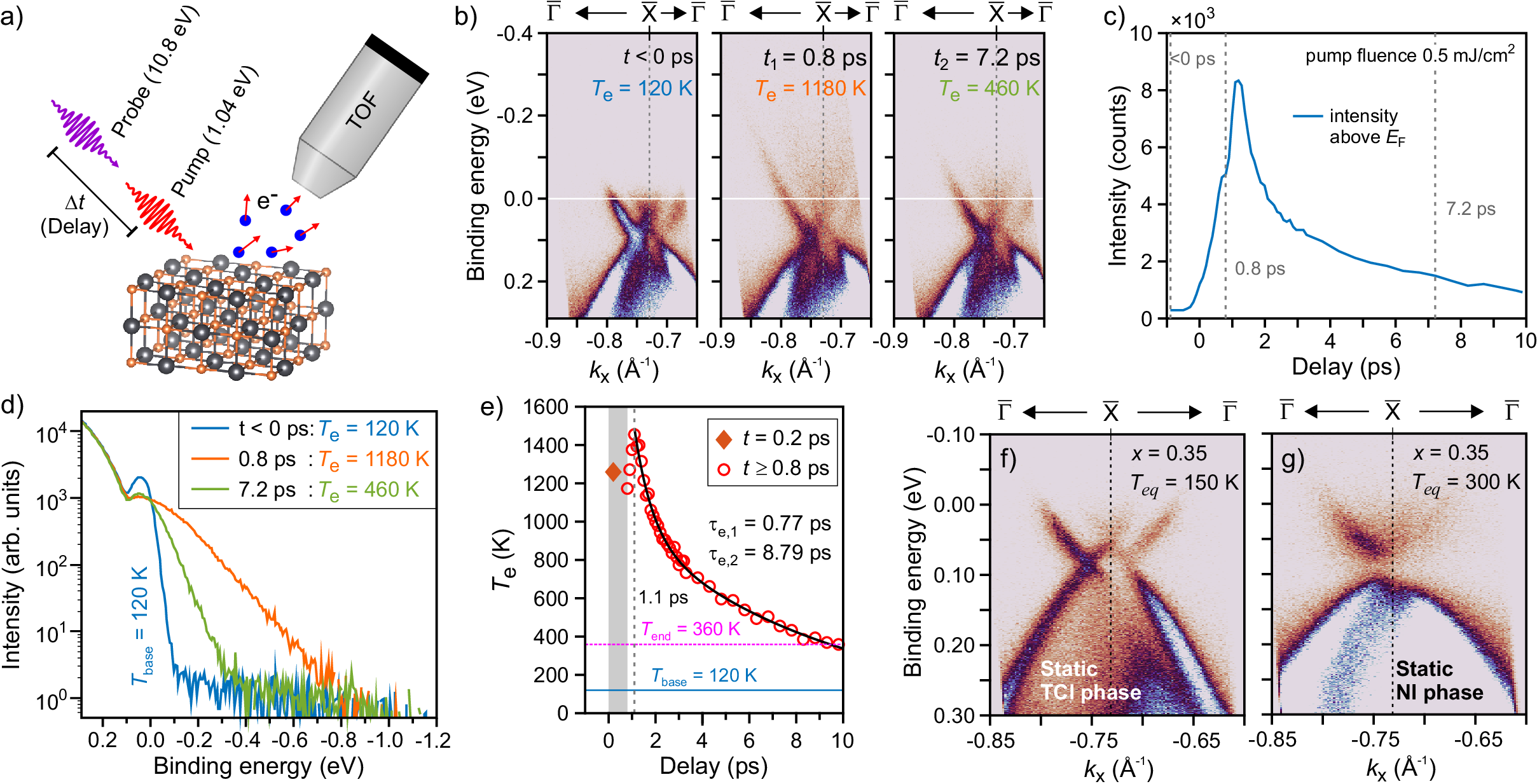}
\caption{\textbf{Electronic response to ultrafast optical excitation.} a) Illustration of the stroboscopic pump-probe ARPES scheme used to measure electron dynamics in Pb$_{0.67}$Sn$_{0.33}$Se. The sample was excited with 1.04~eV photons at a fluence (average) of 0.5 mJ/cm$^2$ and the electronic structure was probed at a variable time delay using 10.8 eV photons. b) $E_\mathrm{b}$--$k$ cuts along \GXG in the SBZ at selected delay times as indicated at the top of each panel. c) The total integrated counts above the Fermi level (0 eV to -1.2 eV) as a function of delay extracted from $E_\mathrm{b}$--$k$ cuts as those displayed in b). Vertical dashed lines show the delays of the cuts in b). d) EDCs extracted from the $E_\mathrm{b}$--$k$ cuts in b) at $k_\mathrm{x} = -0.79$~{\AA}$^{-1}$ plotted using a logarithmic intensity scale. The extracted electronic temperature $T_\mathrm{e}$ for each delay is given in the corresponding legends and also shown in the $E_\mathrm{b}$--$k$ cuts in panel b). e) Fitted $T_\mathrm{e}$ values for the entire measured delay range. Open circles are data points where the electronic population can be described by a Fermi-Dirac distribution around the equilibrium $E_\mathrm{F}$. The gray shaded area, including the diamond data point, indicate the time range where the excited population has not yet reached a Fermi-Dirac distribution around $E_\mathrm{F}$. Values of $T_\mathrm{e}$ before excitation and at the longest measured delay are given by $T_\mathrm{base}$ and $T_\mathrm{end}$, respectively. Solid black line is a fit to the data for $t<1.1$~ps using a double-exponential decay. f) Static ARPES showing the TSS in the TCI phase for a sample with $x=0.35$. g) Static ARPES for a $x=0.35$ sample showing the gap opening in the surface state when raising the sample temperature 40~K above $T_\mathrm{c}$.}
\label{fig2}
\end{figure*}

However, earlier tr-ARPES experiments on Pb$_{0.79}$Sn$_{0.21}$Se ($T_\mathrm{c} = 120$~K) in the TCI phase at $T = 50$~K~\cite{Mogi2024} showed no evidence of a gap opening in the TSS, and thus no phase transition. This raises a key question: does the lattice constant fail to exceed $a_\mathrm{c}$ following optical excitation, or can it transiently surpass $a_\mathrm{c}$ without triggering a transition, implying the involvement of additional mechanisms in the phase transition? To address this, we performed detailed tr-ARPES measurements on Pb$_{0.67}$Sn$_{0.33}$Se at $T = 120$~K ($T_\mathrm{c} = 250$~K). The results are summarized in Fig.~\ref{fig2}.

Figure~\ref{fig2}b displays binding-energy--momentum ($E_\mathrm{b}$--$k$) cuts through the TSS at \Xbar for selected pump–probe delays, aligned with the static data in Fig.~\ref{fig1B}a. The corresponding integrated intensity in a 1.2~eV energy window above the Fermi level ($E_\mathrm{F}$), shown in Fig.~\ref{fig2}c, reveals a strong excitation for positive delays, with the electronic population remaining out of equilibrium for more than 10~ps. Crucially, the TSS remains non-gapped at all delays, as evident from Fig.~\ref{fig2}b.

Energy-distribution curves (EDCs) in the vicinity of \Xbar (Fig.~\ref{fig2}d) show that, after an initial nonthermal regime, the excited electron population can be described by Fermi–Dirac (FD) distributions characterized by an electronic temperature $T_\mathrm{e}$. Starting from $T_\mathrm{base} = 120$~K before excitation, $T_\mathrm{e}$ rises rapidly, reaching $\sim 1500$~K at a delay of 1.1~ps, and subsequently decays following a double-exponential behavior ($\tau_1 = 0.77$~ps, $\tau_2 = 8.79$~ps). At very short delays ($< 0.8$~ps), the electron population cannot be described by a FD distribution (gray shaded region in Fig.~\ref{fig2}e; see \SM), indicating a strongly nonthermal state. For delays between 0.8~ps and 1.1~ps, FD fits are robust but show an increasing $T_\mathrm{e}$ with time. The evolution on this timescale suggests the presence of longer-lived excitations outside the probed energy–momentum window, which gradually relax toward the conduction band minimum at \Xbar. Consequently, the fitted temperature for delays $< 1.1$~ps reflects a local electronic temperature, whereas only at later delays does the system approach a globally thermalized electronic distribution.

Despite $(T_\mathrm{e})_\mathrm{max} \gg T_\mathrm{c} = 250$~K, no gap opening is observed in the TSS (see \SM for further support). For comparison (Figs.~\ref{fig2}f and g), static ARPES measurements on a sample with $x = 0.35$ ($T_\mathrm{c} = 260$~K) show that increasing the lattice temperature by only 40~K above $T_\mathrm{c}$ produces a clear gap in the surface state. These results confirm that optical excitation does not drive the system across the topological phase boundary. In particular, they demonstrate that an elevated electronic temperature alone is insufficient to induce a transition into the NI phase.

\subsection*{Momentum shifts reveal an ultrafast lattice contraction}

\begin{figure*}[t]
\includegraphics[width = 0.7\textwidth]{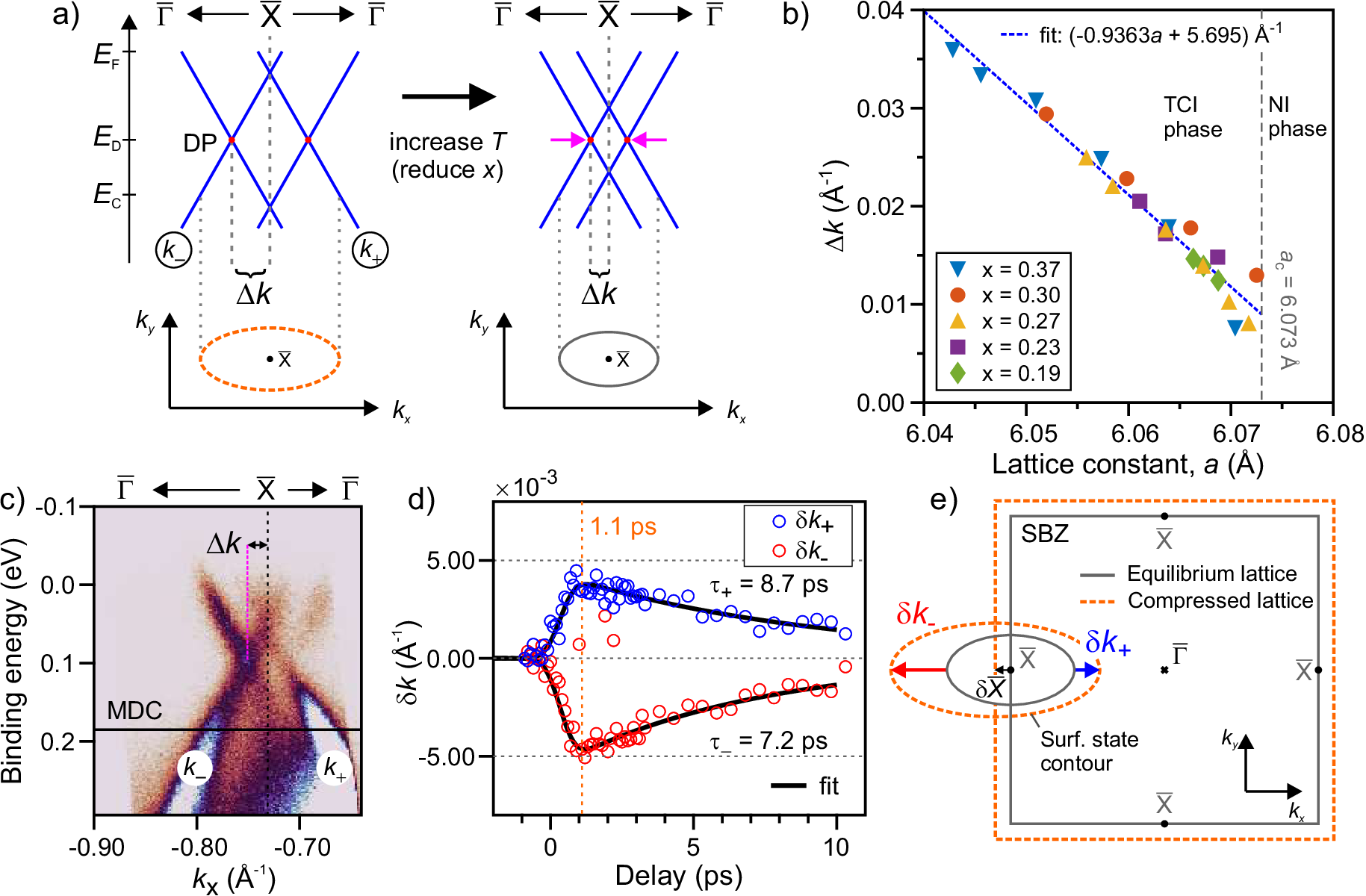}
\caption{\textbf{Changed dispersion of the TSS reveals ultrafast lattice contraction.} a) Illustration of the $E_\mathrm{b}$-$k$ dispersion showing how the momentum separation $\Delta k$ between the Dirac point (at energy $E_\mathrm{D}$) of the TSS and \Xbar changes under equilibrium conditions when the lattice temperature $T$ is increased or $x$ is decreased. The lower part of the sketch illustrates the corresponding change of the constant energy contour at $E_\mathrm{C}$ defined by the $k_{-(+)}$ branches. b) The $x$ and $T$ dependence of $\Delta k$ from ref.~\onlinecite{Wojek2014} replotted as a function of the lattice constant using the $a(x,T)$-relation established by the XRD results (Fig.~\ref{fig1B}e) in the present study. c) Experimental $E_\mathrm{b}$--$k$ cut along \GXG showing the binding-energy position of the MDC that is used to trace the peak positions of the $k_{-(+)}$ branches of the TSS as a function of delay. d) Relative shift ($\delta k$) of the MDC peak positions of the $k_{-(+)}$ branches in panel c) with respect to their equilibrium positions. Solid lines are fits to the data and consists of a Gaussian convoluted with a single-exponential decay function. The decay-time constants are given as $\tau_{-}$ and $\tau_{+}$ for the two curves, respectively. e) Sketch of the constant energy contours from panel a), shown relative to the SBZ and the \Xbar point for each case (not to scale). As the lattice constant differs in the two cases, the size of the SBZs are different. An enlarged SBZ results in an expansion of the contour along with a shift of its center position towards larger momentum. This behavior is consistent with a contraction of the lattice in real space and explains the non-symmetric shift ($\delta k$) seen for the $k_{-(+)}$ branches in panel d).}
\label{fig3}
\end{figure*}

Within a conventional two-temperature model (2TM), energy transferred from the electronic subsystem to the lattice is expected to increase the phonon temperature, leading to thermal expansion. The absence of a TCI–NI transition under optical excitation is therefore unexpected. To assess whether a possible lattice expansion is simply insufficient to increase $a$ beyond $a_\mathrm{c}$, we revisit static ARPES results by Wojek et al.~\cite{Wojek2014}. Figure~\ref{fig3}a illustrates that increasing $T$ or decreasing $x$ reduces the momentum offset $\Delta k$ of the Dirac point (DP) relative to \Xbar. Using the relation between $x$, $T$, and $a$ established by our XRD measurements, we replot $\Delta k$ as a function of the lattice constant. As shown in Fig.~\ref{fig3}b, all data collapse onto a universal linear dependence, indicating that $\Delta k$ serves as a direct proxy for the lattice constant.

If optical excitation were to induce lattice expansion, the corresponding increase in $a$ would lead to a reduction of $\Delta k$, i.e., a shift of the Dirac points toward \Xbar. However, momentum-distribution-curve (MDC) analysis of the $k_{-}$ and $k_{+}$ branches of the TSS (Fig.~\ref{fig3}c) reveals the opposite behavior: both branches shift away from \Xbar following excitation. The extracted momentum shifts $\delta k_{-(+)}$ (Fig.~\ref{fig3}d) reach their maximum at $\sim 1.1$~ps, coinciding with the peak electronic temperature, and decay on a timescale comparable to the slower component of electronic cooling. Notably, the shifts are asymmetric, with $|\delta k_{-}| > |\delta k_{+}|$.

Because the cut in Fig.~\ref{fig3}c follows the \GXG high-symmetry line, the momentum distances to \Xbar for the $k_{-}$ and $k_{+}$ branches are expected to be equal. The observed asymmetry in $\delta k_{-(+)}$ therefore indicates that \Xbar itself shifts toward larger momentum, corresponding to an expansion of the SBZ and, consequently, a contraction of the lattice. This interpretation is consistent with Fig.~\ref{fig3}b, where the increase in $\Delta k$ -- corresponding to a larger separation between the $k_{-}$ and $k_{+}$ branches -- likewise indicates a reduced lattice constant. Figure~\ref{fig3}e schematically illustrates how an isotropic lattice contraction leads to an expansion of the SBZ and a corresponding shift (and expansion) of the constant-energy contour, providing a consistent explanation of the observed momentum shifts.

From the perspective of a conventional two-temperature model, such a lattice contraction is unexpected. Although phonon heating is anticipated on the relevant timescales, the absence of the expected thermal expansion indicates that the lattice response is dominated by a competing nonthermal mechanism.

\section{Discussion}

The XRD results in Fig.~\ref{fig1B} demonstrate that the $x$-–$T$ dependence of the \PbSnSe topological phase diagram is governed primarily by the lattice constant: increasing $x$ has the same effect on $a$ as lowering $T$. This equivalence is further reflected in the electronic structure of the TSS, where $\Delta k$ in Fig.~\ref{fig3}b is uniquely determined by the lattice constant. Together, these findings indicate that crystal-field splitting plays a central role in driving the inversion of the bulk band gap as the lattice constant crosses the critical value $a_\mathrm{c}$. Moreover, $\Delta k$ provides an experimentally accessible ARPES proxy for tracking the lattice constant.

\subsection*{Quantifying the pump-induced lattice contraction}

For the $x = 0.33$ sample studied by tr-ARPES, we estimate the maximum pump-induced change in the lattice constant $\Delta a_\mathrm{max}$ from fits to $\delta k_{-}$ and $\delta k_{+}$ (Fig.~\ref{fig3}d). Assuming that the Dirac point at $k = -0.76$~\AA$^{-1}$ in Fig.~\ref{fig3}c shifts by the same amount as the $k_{-}$ branch intersected by the MDC ($\delta k_\mathrm{max} = 0.0042$~\AA$^{-1}$), Fig.~\ref{fig3}b yields $\Delta a_\mathrm{max} \approx -0.45$~pm. 

Alternatively, using the relative shift of \Xbar ($\delta \bar{X}$) extracted from the fits to $\delta k_{-}$ and $\delta k_{+}$, and the relation $a = \sqrt{2}\pi/k_{\bar{X}}$, where $k_{\bar{X}}$ is the momentum position of \Xbar, we obtain $\Delta a_\mathrm{max} \approx -0.35$~pm. The close agreement between these independent estimates (see \SM for details) confirms that the lattice response is indeed a contraction. This also highlights the sensitivity of $\Delta k$ as a probe of sub-picometer lattice changes. For comparison, a thermal contraction of $\Delta a = -0.40$~pm would correspond to a temperature change of $\Delta T = -33$~K at $T = 120$~K.

The schematic in Fig.~\ref{fig3}e illustrates how an isotropic (three-dimensional) lattice contraction modifies the SBZ and the constant-energy contour of the TSS. The asymmetry between $\delta k_{-}$ and $\delta k_{+}$ arises naturally from the combined effect of an expansion of the TSS contour and a shift of \Xbar due to the enlarged SBZ. The ARPES data directly reveal a reduction of the in-plane lattice constant consistent with this picture, although without direct sensitivity to the out-of-plane component. Previous reports of optically induced lattice contraction have largely focused on thin films and layered materials \cite{Chen2011,Mannebach2017,Kim2021,Kumazoe2018,Park2024}, where the structural response is predominantly out-of-plane and attributed to blast-force-like acoustic waves \cite{Chen2011} or pump-induced changes in interlayer van der Waals interactions \cite{Mannebach2017}. The observation of an in-plane contraction in the present three-dimensional crystal therefore points to a different underlying mechanism.

\subsection*{Band inversion enables photoinduced lattice contraction}

We propose that this mechanism originates from the band inversion in the topological phase, which alters the response of the chemical bonding to electronic excitation. In particular, optical excitation in the TCI phase strengthens the interatomic bonds, leading to a transient contraction of the lattice that is likely three-dimensional in nature. In the following, we examine this hypothesis in more detail.

First, tight-binding (TB) calculations for the related compound (Pb,Sn)Te support the interpretation of an isotropic lattice change. Using the Sn concentration ($x$) as a proxy for the lattice constant, we reproduce the observed asymmetric momentum shifts of the $k_{-}$ and $k_{+}$ branches by assigning a specific value of $a$ to each pump–probe delay. This directly links the measured electronic band-structure dynamics to a transient lattice contraction. Additional details are provided in the \SM.

\begin{figure*}[t]
\includegraphics[width=\textwidth]{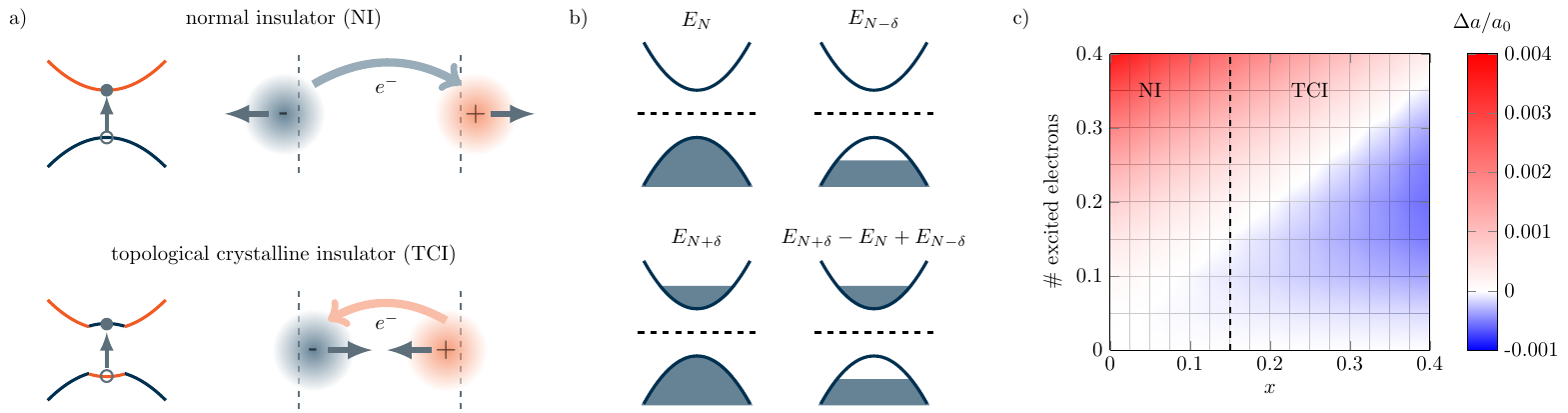}
\caption{\textbf{Explaining lattice contraction using ab initio calculations.} a) Schematic comparison of the dispersion relations of the normal insulator phase and the topological crystalline insulator phase. Exciting electrons from the valence to the conduction band will, after initial relaxation processes, lead to a depletion of carriers in the top of the valence band and accumulation in the bottom of the conduction band. The associated charge transfer from negatively charged anions to positively charged cations in the normal insulator phase weakens the atomic bonding. The behavior is inverted in the topological crystalline insulator phase leading to a charge transfer from cations to anions and a bond strengthening. b) Illustration showing the approximation of a total energy calculation for a non-thermal electron filling after optical excitation of $\delta$ electrons per unit cell, by comparing total energies for unit cells filled with $N$, $N-\delta$, and $N+\delta$ electrons. c) Calculated relative change of the lattice constant as a function of the number of excited electrons per unit cell and composition $x$. Regions of lattice expansion (contraction) are shown in red (blue). The vertical dashed line guides the eye, separating the NI and TCI phases for $\delta = 0$.}
\label{fig4:dft}
\end{figure*}

For $x < 0.40$, \PbSnSe crystallizes in a rock-salt structure consisting of cationic Pb/Sn and anionic Se atoms, such that the conduction and valence bands have distinct orbital character (Fig.~\ref{fig4:dft}a). In the NI phase, the valence-band maximum is primarily anionic and the conduction-band minimum cationic. Optical excitation therefore transfers electrons from anions to cations, reducing the electrostatic attraction between the sublattices and weakening the interatomic bonds. In the TCI phase, band inversion reverses this orbital character: the valence-band maximum becomes cationic, while the conduction-band minimum becomes anionic (see Ref.~\onlinecite{Dziawa2012} and \SM). Optical excitation then leads to charge transfer from cations to anions, strengthening the interatomic bonding and driving a transient lattice contraction.

The bonding in \PbSnSe has been described as metavalent \cite{Raty2019}, intermediate between ionic and covalent bonding. Within this framework, the same mechanism can be understood as a redistribution of carriers between bonding and antibonding states. In the band-inverted phase, photoexcitation increases the occupation of bonding states while simultaneously depopulating antibonding states, resulting in a transient strengthening of the atomic bonds rather than the bond weakening typically observed in conventional semiconductors.

To verify this picture, we performed ab initio calculations of the lattice constant under non-thermal electron occupation (Fig.~\ref{fig4:dft}b and Methods). These calculations are agnostic with respect to the nature of bonding and provide a qualitative explanation of the observed contraction. As shown in Fig.~\ref{fig4:dft}c, the NI phase ($x \lesssim 0.15$) exhibits a lattice expansion (up to $\sim 0.4\%$) upon excitation, whereas the TCI phase ($x > 0.15$) shows a contraction of up to $\sim 0.1\%$ for moderate excitation densities. A crossover from contraction to expansion occurs at higher excitation densities, consistent with the competition between bonding and antibonding contributions. While the exact crossover depends on carrier density and doping, the qualitative behavior remains robust.

The electronic origin of the lattice contraction is further supported by the temporal correlation between the contraction and the electronic excitation, both of which peak at $\sim 1.1$~ps and decay on comparable timescales. While phonon heating is expected to follow the electronic excitation, the absence of a detectable thermal expansion within the measured time window indicates that the nonthermal contraction dominates the lattice response. At longer delays, electron–phonon equilibration is expected to lead to thermal expansion. Our measurements do not extend to sufficiently long times to capture this crossover, but future tr-ARPES studies at extended delays could reveal this transition. In addition, investigating the fluence dependence of the lattice response may further test the predicted crossover between contraction and expansion in the TCI phase, providing a pathway toward ultrafast optical control of topological order in \PbSnSe.

In summary, we establish the lattice constant as the key control parameter governing the topological-to-normal-insulator transition in \PbSnSe and determine a critical value of $a_\mathrm{c}=6.073$~\AA. We further demonstrate a linear relationship between $\Delta k$, the momentum distance from the Dirac point to \Xbar, and the lattice constant. Tracking $\Delta k$ following optical excitation using tr-ARPES reveals that a sub-picosecond lattice contraction dominates the structural response. Rather than increasing the lattice constant toward the critical value $a_\mathrm{c}$, optical excitation instead induces a contraction that drives the system further into the topological phase. This provides a direct explanation for the absence of an optically driven TCI--NI phase transition in both our measurements and previous tr-ARPES studies. Total-energy calculations show that this behavior originates from the band inversion itself, which reverses the character of the photoinduced charge transfer and leads to a transient strengthening of the atomic bonds. More broadly, the present results identify band inversion as a mechanism that fundamentally alters the nonequilibrium lattice response to optical excitation.

\section{Methods}
\subsection*{Crystal growth}

Single crystals of \PbSnSe with varying Sn concentrations were grown using the self-selecting vapor growth method. The crystals were characterized by X-ray diffraction (XRD) and energy-dispersive X-ray spectroscopy (EDX) to verify their structural quality and Sn content. All crystals used for the XRD and ARPES measurements in this study originate from the same batch as those used in refs.~\onlinecite{Dziawa2012,Wojek2014,Wojek2015}.

\subsection*{X-ray diffraction measurements and analysis}

Temperature-dependent XRD measurements were performed using a Rigaku diffractometer at Uppsala University, Sweden, with Cu K$\alpha_1$ radiation ($\lambda = 1.540598$~Å). \PbSnSe single crystals with Sn contents $x = 0.00$, 0.19, 0.27, 0.33, and 0.37 were examined. Diffractograms were collected at 50~K intervals over the temperature range 100~K to 300~K. The diffraction data were integrated into powder-like diffractograms and background-subtracted using the standard procedures implemented in the Rigaku CrysAlisPro software package. For each $x$ and $T$, the lattice constant $a$ was determined as the average of values extracted from fits to the (044), (066), (026), and (226) Bragg reflections. The positive (negative) error bars shown in Fig.~\ref{fig1B}e correspond to $a_\mathrm{max} - a_\mathrm{mean}$ ($a_\mathrm{min} - a_\mathrm{mean}$) and are smaller than the data markers.

The $x$–-$T$ dependence of $a$ was modeled using the empirical two-dimensional function
\begin{multline}
    a(x,T) = \hat{a}(T)\cdot \check{a}(x,T) \cdot a(x) \\ =\frac{(c_1+c_2T+c_3T^2+c_4T^3+c_5T^4)}{c_6}\cdot \left(1-\frac{(c_7-T)}{c_8}\cdot \frac{x}{c_9}\right) \cdot(c_6+c_{10}x),
\end{multline}
where the constants have the following values:
$c_1 = 6.091$~{\AA}, $c_2 = 1.01723 \times 10^{-5}$~{\AA}K$^{-1}$, $c_3 = 8.36225 \times 10^{-7}$~{\AA}K$^{-2}$ , $c_4 = -2.8424 \times 10^{-9}$~{\AA}K$^{-3}$, $c_5 = 3.50792 \times 10^{-12}$~{\AA}K$^{-4}$, $c_6 = 6.1211$~{\AA}, $c_7 = 300$~K, $c_8 = 3.8 \times 10^5$~K, $c_9 = 0.37$, and $c_{10} = -0.1187$~{\AA}.

This function combines the room temperature $x$-dependence of the lattice constant, $a(x)$, obtained from our XRD data with a temperature-scaling factor, $\hat{a}(T)$, based on a fit to the temperature dependence of the PbSe lattice constant from ref.~\onlinecite{Preier1979}. An additional scaling factor, $\check{a}(x,T)$, accounts for the slightly larger temperature derivative of $a$ observed for higher $x$. Compared to the results obtained in ref.~\onlinecite{Preier1979}, small adjustments to the modeled lattice constant of PbSe at $T = 0$~K and 300~K were made to better match the present experimental data.

\subsection*{Static and time-resolved photoemission}

Static angle-resolved photoemission spectroscopy (ARPES) measurements on the $x = 0.35$ sample (Figs.~\ref{fig2}f–g) were performed using 10.5~eV photons and the high energy-resolution ARPES setup at the Baltazar laser-ARPES laboratory, KTH Royal Institute of Technology, Sweden~\cite{Berntsen2011}. The energy and momentum resolutions were better than 5~meV and 0.001~\AA$^{-1}$, respectively, and the repetition rate of the light source was 500~kHz. Sample temperature was controlled by a closed-cycle cryostat (Advanced Research Systems) equipped with a calibrated Si diode sensor. Data were collected around the \Xbar point in the surface Brillouin zone at sample temperatures of 150~K and 300~K.

Time-resolved ARPES (tr-ARPES) measurements were conducted on a $x = 0.33$ sample at an equilibrium temperature of 120~K using the tr-ARPES setup at the Baltazar laser-ARPES laboratory at KTH Royal Institute of Technology, Sweden~\cite{Guo2022}, operated at a 250~kHz repetition rate. Pump pulses at 1200~nm wavelength ($1.04$~eV) with $ <100$~fs duration were generated by an optical parametric amplifier (Twinstarzz, Fastlite), and the 10.8~eV harmonic from the high-harmonic generation (HHG) light source served as the probe. The combined energy, temporal, and momentum resolutions were 10~meV, 250~fs, and $< 0.001$~\AA$^{-1}$, respectively. The incident pump fluence, averaged over the illuminated surface area, was $\sim 0.5$~mJ\xspace cm$^{-2}$.

Two tr-ARPES datasets were acquired: (i) a high-statistics measurement at four selected pump–probe delays ($t < 0$~ps, 0.2~ps, 0.8~ps, and 7.2~ps) and (ii) a discrete delay scan from $-1$~ps to $+10$~ps, with 100~fs steps for $-1\, \mathrm{ps} \leq t \leq 3\, \mathrm{ps}$ and 500~fs steps for $3\, \mathrm{ps} < t \leq 10\, \mathrm{ps}$. The ARPES intensity slices shown in Figs.~\ref{fig1B}a-c, \ref{fig2}b, and \ref{fig3}c, as well as the EDCs in Fig.~\ref{fig2}d, were extracted from the high-statistics dataset, while data in Figs.~\ref{fig2}c, \ref{fig2}e, and \ref{fig3}d were taken from the delay-scan dataset.

\subsection*{Analysis of tr-ARPES data}

Time zero ($t = 0$~ps) was defined as the midpoint of the rising edge of the integrated photoemission intensity within a binding-energy interval $E_b \in \left[-1.2, -0.65 \right]$~eV. The electronic temperature $T_\mathrm{e}$ at each delay was determined by extracting energy-distribution curves, $I_\mathrm{EDC}$, from $E_\mathrm{b}$--$k$ slices (such as those in Fig.~\ref{fig2}b, integrated over a 0.14~{\AA}$^{-1}$ wide window in $k_\mathrm{x}$ and $k_\mathrm{y}$ centered at $k_{\parallel} = \left(-0.79, 0\right)$~{\AA}$^{-1}$.

Taking the logarithm of $I_\mathrm{EDC}$ and assuming $(E_\mathrm{F}-E_\mathrm{b}) \gg k_\mathrm{B} T_{e}$, we use the approximation $\ln(I_\mathrm{EDC}) \approx \left(E_b - E_\mathrm{F}\right) / (k_\mathrm{B} T_{e})$, which allowed extraction of $T_{e}$ from linear fits to the logarithmic intensity in a region sufficiently far above $E_\mathrm{F}$. This approach provides a robust estimate of the electronic temperature, as it primarily probes excited states in the bulk conduction band.

Momentum-distribution curves (MDCs) used to analyze the time evolution of the $k_{-}$ and $k_{+}$ branches (Figs.~\ref{fig3}c)–d)) were extracted at $E_\mathrm{b} = 0.190$~eV, integrating over $\Delta E_\mathrm{b} \approx 50$~meV and 0.024~{\AA}$^{-1}$ along $k_\mathrm{y}$. Lorentzian fits were used to determine the momentum positions of $k_{-}$ and $k_{+}$ at each delay.

\subsection*{Tight-binding calculations}
We employ the relativistic empirical tight-binding (TB) model from Lent et al.~\cite{Lent1986}, based on the nearest-neighbor formulation introduced by Slater and Koster~\cite{Slater1954}, and solve the corresponding eigenvalue problem for a slab geometry comprising 280 unit cells along the out-of-plane direction. The model is parameterized for \PbSnTe, with TB parameters for intermediate Sn concentrations ($x$) obtained by interpolating between those of PbTe and SnTe within the virtual-crystal approximation (VCA).

At a qualitative level, the model successfully captures the composition dependence of the band structure in \PbSnSe, despite being parameterized for \PbSnTe. Quantitative differences remain, notably in the critical composition and the size of the surface Brillouin zone, the latter which can be attributed to the differing lattice constants of the two compounds.\\

\subsection*{\textit{Ab initio} crystal lattice optimization}
The lattice constant for a non-thermal electron distribution in Pb$_{1-x}$Sn$_x$Se was obtained from ab initio total energy calculations, in the framework of the density functional theory using the Vienna Ab initio Simulation Package (VASP)~\cite{vasp}. The exchange-correlation functional was approximated using the generalized gradient approximation of Perdew, Burke and Ernzerhof~\cite{perdew1996generalized}. The used energy cut-off was set to 240~eV. The varying Sn content in Pb$_{1-x}$Sn$_x$Se was modeled using the virtual crystal approximation (VCA)~\cite{vca1,vca2}, which reproduces the linear decrease of the lattice constant with increasing Sn content $x$ (see \SM Fig.~S8a). In each calculation we used a $14\times14\times14$ $\vec{k}$-point mesh.

For each Sn content $x$, we computed the total energy for varying lattice constant $a$ and electron excitation $\delta$. To approximate the non-thermal electron distribution after optical excitation, we used the scheme shown in Fig.~\ref{fig4:dft}b. Performing a self-consistent reference calculation for the energy $E_N(a)$, gives the ground state charge distribution and, therefore, fixing the orbital ordering. Here, $N=20$ is the number of valence electrons per primitive unit cell. We then performed non-self-consistent calculations for varying number of electrons $\delta$, giving $E_{N+\delta}(a)$. The total energy after optical excitation was approximated using $E_{tot}(a) = E_{N+\delta}(a) - E_N(a) + E_{N-\delta}(a)$ (Fig.~\ref{fig4:dft}b). The non-equilibrium lattice constant follows by minimizing $E_{tot}(a)$ (see \SM Fig.~S8b). 

Note that the calculations assume a Fermi level positioned at mid-gap in equilibrium. Experimentally, the measured sample is $n$-type doped, which lowers the number of excited electrons required to achieve a particular lattice contraction or expansion. Nevertheless, the qualitative behavior shown in Fig.~\ref{fig4:dft}c remains unchanged.

\bibliography{references_main.bib}

\section{Acknowledgements}
We thank A.~Szczerbakow and T.~Story from the Institute of Physics, Polish Academy of Sciences, for providing the samples used in this study. This work has received financial support from the Swedish Research Council (2019-03486), and the Knut and Alice Wallenberg Foundation (2018.0104). A.G.C. acknowledges the financial support of the Zernike Institute for Advanced Materials. M.D. acknowledges financial support from the Göran Gustafsson Foundation, Swedish Research Council under Grant No: 2022-03813 and the Carl Trygger Foundation. R.M.G. acknowledges support from the Knut and Alice Wallenberg Foundation (Grant No. 2023.0087), the Swedish Research Council (VR starting Grant No. 2022-03350), the Olle Engkvist Foundation (Grant No. 229-0443), and Chalmers University of Technology, via the Department of Physics and the Areas of Advance Nano and Materials Science. J.W. acknowledges the Swedish Research Council for funding via Grant No. 2021-04379. J.W. was partially supported by the Wallenberg Initiative Materials Science for 
Sustainability (WISE) funded by the Knut and Alice Wallenberg Foundation 
(KAW).

 \section{Author contributions}
 M.H.B. and O.T. planned the experiment. M.H.B., W.C., A.G.C., M.D., Q.G., and C.L. performed the tr-ARPES experiment. Temperature-dependent X-ray diffraction measurements were carried out by W.C., D.P., A.O., J.Wu, and M.H.B. The tight-binding calculations were performed by W.C., M.H.B., and O.T., while the DFT and total-energy calculations were conducted by R.M.G. M.H.B. analyzed the tr-ARPES data, and W.C. and M.H.B. analyzed the X-ray diffraction data. M.H.B., W.C., J.Wu, J.W, R.M.G., and O.T. interpreted the results. M.H.B. wrote the manuscript in collaboration with W.C. and O.T. All authors commented on the manuscript.

 \section{Competing interests}
The authors declare no competing interests.

\end{document}


\title{Optically driven ultrafast lattice contraction in a topological crystalline insulator - Supplementary information}

\author{M.~H.~Berntsen}
\thanks{Corresponding authors: mhbe@kth.se, oscar@kth.se}
\affiliation{\KTH}
\author{W.~Chen}
\affiliation{\KTH}
\author{D.~Phuyal}
\affiliation{\KTH}
\author{J.~Wu}
\affiliation{\KTH}
\author{A.~Grubi\v{s}i\'{c}-\v{C}abo}
\affiliation{\KTH}
\affiliation{\Groningen}
\author{M.~Dendzik}
\affiliation{\KTH}
\author{Q.~Guo}
\affiliation{\KTH}
\author{A.~Orthaber}
\affiliation{\UU}
\author{C.~Li}
\affiliation{\KTH}
\author{R.~M.~Geilhufe}
\affiliation{\Chalmers}
\author{J.~Weissenrieder}
\affiliation{\KTH}
\author{O.~Tjernberg}
\thanks{Corresponding authors: mhbe@kth.se, oscar@kth.se}
\affiliation{\KTH}


\maketitle


\section{Electronic structure and temperature in sub-picosecond delay range}

\begin{figure*}[t]
\includegraphics[width=0.8\textwidth]{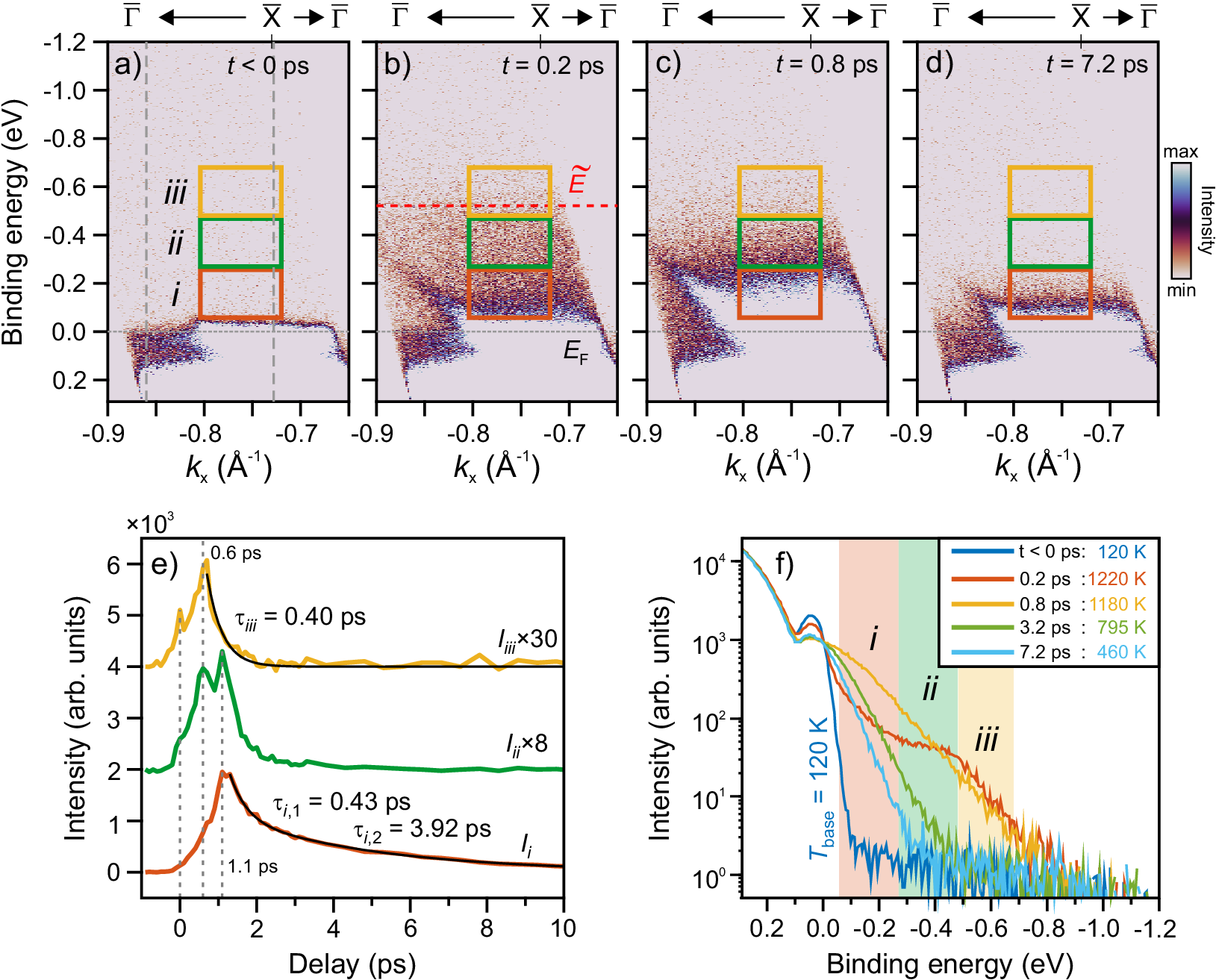}
\caption{\textbf{Sub-picosecond dynamics.} a)-d) Band structure along the $\bar{\Gamma}$-$\bar{X}$-$\bar{\Gamma}$ direction for a sample with $x=0.33$ measured at $T=120$~K. Colormap is saturated to enhance the excitation above $E_\mathrm{F}$. e) Delay curves showing integrated intensity in the energy and momentum regions $i$ (yellow), $ii$ (green) and $iii$ (reg) shown in panels a)-d). f) Energy-distribution curves integrated over the momentum range between the two vertical (dashed) lines in panel a). Energy regions $i$--$iii$ are marked for clarity and the fitted electronic temperature for the different delays are given in the corresponding legends.}
\label{fig1SI}
\end{figure*}

Figure~\ref{fig1SI}a)-d) shows the measured band structure along the \GXbar direction for given delays. The colormap has been saturated to increase visibility of the excited intensity above $E_\mathrm{F}$. The integrated intensities in the three regions,  $i$ to $iii$, are displayed as a function of delay in panel e). The delay curve in panel e) for region $iii$ (yellow) shows a distinct peak at $t=0$~ps, which we have used to define $t_0$ in our experiment. The subsequent decrease in intensity, followed by another rise, shows that this energy interval is being filled by carriers that relax from higher excited states with a delay of approximately 0.6 ps. This could suggest that there are states at higher energy with increased lifetime. A similar multi-peak delay curve is observed for region $ii$ (green) with the initial excitation peak at $t=0$ ps seen as a weak hump, followed by two distinct peaks at 0.6 ps and 1.1 ps, respectively. The peak at 0.6 ps coincides with the maximum observed in the delay curve for region $iii$. The same peaks are also observed in the delay curve for region $i$ (red), although for this energy range the global maximum in integrated intensity occurs at 1.1 ps delay.

What the delay curves in panel e) tells us is that there are processes occurring in the system on a sub-picosecond time scale that delay the formation of a thermalized electronic population. To better visualize what is happening on this time scale, we plot the integrated intensity in the momentum range given by the two vertical dashed lines in panel a) as a function of energy. The results are shown in panel f). By using a logarithmic intensity scale a Fermi-Dirac distributed electronic population will give rise to a linear region in an energy range sufficiently above $E_\mathrm{F}$ where the electronic temperature $T_\mathrm{e}$ can be extracted from the linear slope. We note that for the curve representing a delay of 0.2 ps the intensity distribution shows a linear behavior in region $iii$ that is different from how it behaves in regions $i$ and $ii$. Comparing with the corresponding curves extracted before $t_0$ and at longer delays ($t\geq 0.8$~ps) it becomes clear that the electronic population follows a Fermi-Dirac distribution with respect to the Fermi level ($E_\mathrm{b} = 0$~eV) for the longer delays. At $t=0.2$~ps, however, the excited electrons are instead described by a Fermi–Dirac distribution centered around a transiently elevated level $\tilde{E}_\mathrm{F}$ (see dashed red line in panel b)). This transient distribution gives rise to the peaks observed at 0.6 ps in the delay traces shown in panel e), and indicates that the electronic system has not yet fully equilibrated. Instead, electrons from higher-energy states outside the probed energy–momentum window continue to relax into the monitored region.

\section{Absence of energy gap opening at Dirac point after optical excitation}

\begin{figure*}[t]
\includegraphics[width = 0.9\textwidth]{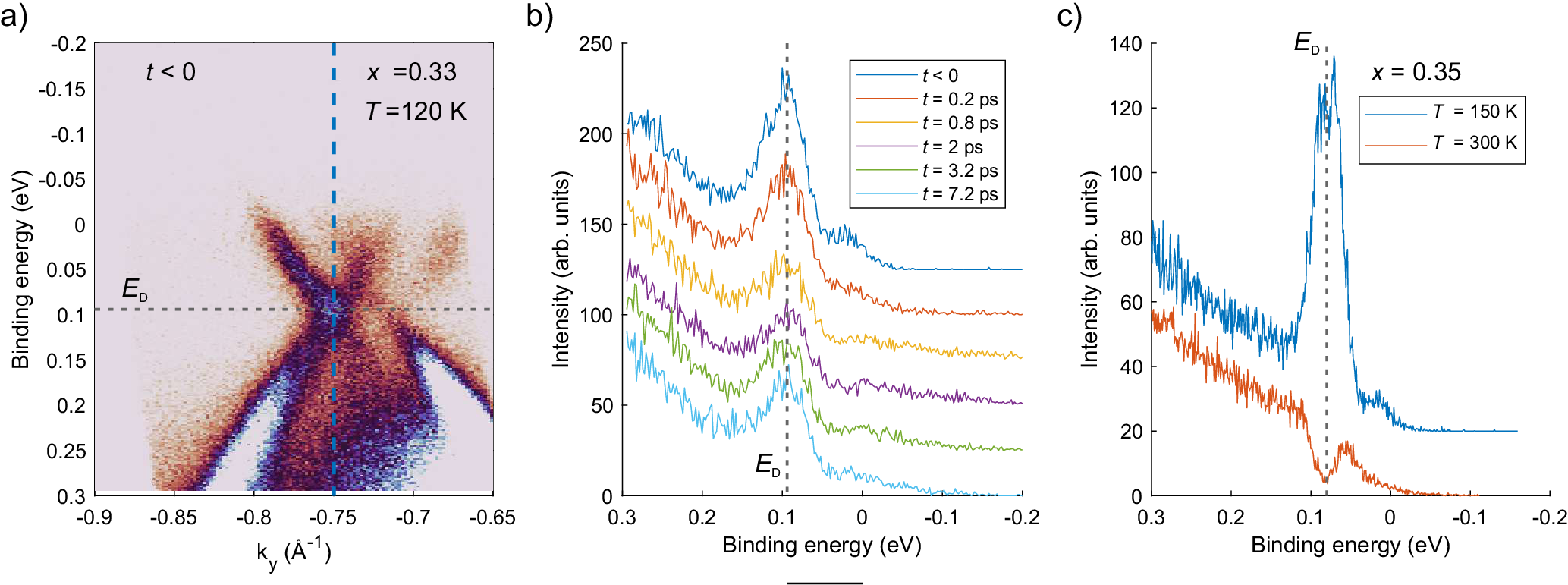}
\caption{\textbf{Demonstrating no energy gap opening at the Dirac point.} a) Static band dispersion of the surface state along $\Gamma$-$\bar{X}$-$\Gamma$ direction in the $x=0.33$ sample. $E_\mathrm{D}$ marks the binding energy of the Dirac point. b) EDCs from the $x=0.33$ sample at the location marked by the dashed line in a) for different pump-probe delay times. c) EDCs at equilibrium temperatures 150 K and 300 K taken through the Dirac point for a sample with $x=0.35$, having a $T_\mathrm{c} = 260$~K.}
\label{fig3SI}
\end{figure*}

Figure~\ref{fig3SI} investigates the detailed energy spectrum around the Dirac point of the surface state for different delays. The energy-distribution curves (EDCs) plotted in panel b) are extracted at the momentum location indicated by the vertical dashed line (blue) in panel a). As apparent from the EDCs, no obvious gap opening is observed at the Dirac point (at $E_\mathrm{b} \simeq 0.1$~eV) -- in spite of the electronic temperature far exceeding the critical temperature for the TCI to NI phase transition, as reported in Fig.~3e) in the main text. As a reference, panel c) shows similar EDCs from another sample ($x = 0.35$) at equilibrium conditions and temperatures below and above $T_c$. The sample with $x=0.35$ has a $T_c = 260$~K and the EDC corresponding to the 300~K data in panel c) demonstrates the stark difference in the energy spectrum when exceeding $T_c$ with as little as 40~K.

\section{Determining dynamic change in lattice constant}

\begin{figure*}
\includegraphics[width=0.9\textwidth]{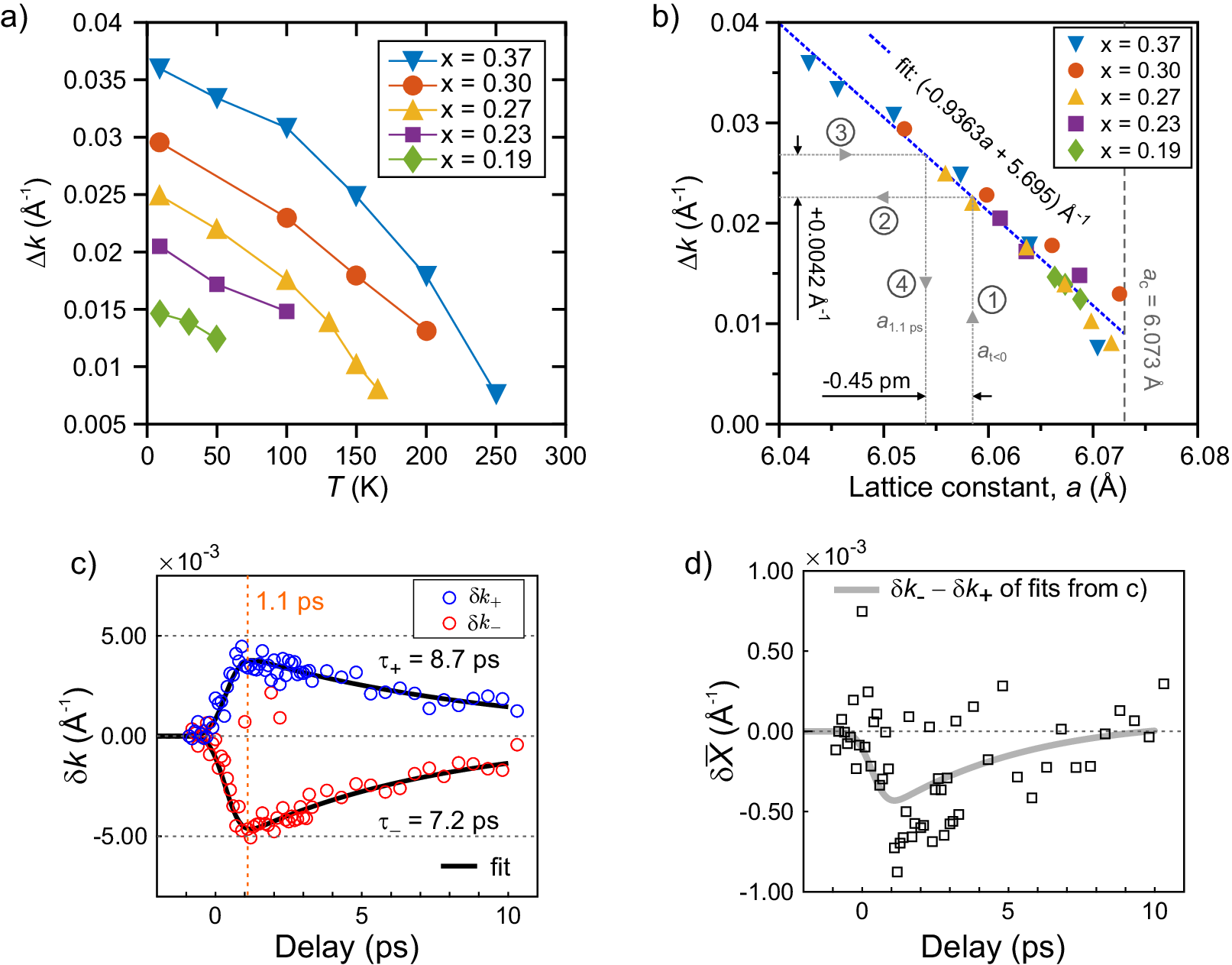}
\caption{\textbf{Change in lattice constant derived from changes in band structure and change in size of the surface Brillouin zone.} a) Data from ref.~\cite{Wojek2014} replotted. $\Delta k$ represents the momentum separation between the Dirac point of the surface state and the \Xbar point in the SBZ. b) Transforming the data in panel a) by using the composition and temperature dependence of the lattice constant, $a(x,T)$, as determined by XRD in the present study. c) Relative momentum shifts $\delta k_{-}$ and $\delta k_{+}$ of the $k_{-}$ and $k_{+}$ branches of the surface state, respectively, with delay. d) Relative shift of the \Xbar point in the SBZ with delay.}
\label{fig2SI}
\end{figure*}

Figure~\ref{fig2SI} demonstrates two different approaches to determining the maximum change in lattice constant ($\Delta a_\mathrm{max}$) caused by the optical excitation of the Pb$_{0.67}$Sn$_{0.33}$Se sample. Panel a) re-plots the results of Wojek et al.~\cite{Wojek2014} showing how the momentum separation between the \Xbar point and the Dirac point of the surface state varies for different compositions ($x$) with temperature ($T$). In panel b) we take the data from a) and use the fact that specific values of $x$ and $T$ can be mapped to a value of the lattice constant $a$ (as deduced from our XRD data presented in Fig.~2e) in the main text) to plot the momentum separation $\Delta k$ of the Dirac point from the \Xbar point as a function of $a$, as shown in Fig.~\ref{fig2SI}b). The linear relation indicates a general trend in this system, supporting the notion of the lattice constant being the dominant tuning parameter also for $\Delta k$. 
Panel c) displays the relative momentum shift of the two branches of the surface state intersected by the MDC shown in Fig.~4c) in the main text. By assuming that the Dirac point will experience the same momentum shift as the part of the branch that is monitored in c) we use the average maximum magnitude of the fitted $\delta k_{-}$ and $\delta k_{+}$ curves (Fig.~\ref{fig2SI}c) to estimate $\Delta k_\mathrm{max}$. We get $\Delta k_\mathrm{max} = 0.0042$~\AA$^{-1}$ which we together with panel b) can use to read out a change in lattice constant of $\Delta a_\mathrm{max} = -0.45$~pm.

Since the momentum location of the \Xbar point is given by the the midpoint between the $k_{-}$ and $k_{+}$ branches we can use their extracted peak positions to determine how \Xbar shifts in time after the excitation. Panel d) shows data (squares) and a fit (solid curve, calculated from fits of $\delta k_{-}$ and $\delta k_{+}$ in c)) of the relative momentum position of \Xbar with delay. From the fitted curve in d) we extract a maximum deviation from equilibrium of $\delta \bar{X}_\mathrm{max} = -0.42\cdot 10^{-3}$~\AA$^{-1}$ for the \Xbar point. We now use the direct lattice vectors, with the equilibrium lattice constant $a_0 = 6.0585$~\AA \xspace(at $T = 120$~K) to calculate the reciprocal lattice vectors and to express the expected momentum location of the \Xbar point. For the face-centered cubic direct lattice, \Xbar in momentum space is located at a distance $\sqrt{2}\pi/a_0$~\AA$^{-1}$ from $\bar{\Gamma}$. The experimentally determined $\delta \bar{X}_\mathrm{max}$ can now be used to estimate the maximum change in $a$ through the relation $\Delta a_\mathrm{max} = \frac{\sqrt{2}\pi}{(\sqrt{2}\pi/a_0 - \delta \bar{X}_\mathrm{max})} - a_0$, yielding $\Delta a_\mathrm{max} = -0.35$~pm.
The two approaches yield similar values for $\Delta a_\mathrm{max}$ ($<$30\% difference). However, based on panel b), we expect that the method based on $\delta \bar{X}$ under estimates $\Delta a_\mathrm{max}$ due to the difficulty of the fit to describe the noisy data.

\section{Ruling out rigid energy shift of band structure}

An alternative explanation for the increased momentum separation between the $k_{-}$ and $k_{+}$ branches is a rigid shift of the band structure toward lower binding energy induced by a surface photovoltage (SPV)~\cite{Boschini2024}. Such an energy shift would cause the MDC in Fig.~4c) in the main text to intersect the topological surface state at an effectively higher binding energy. The surface photovoltage (SPV) effect can frequently be observed in time-resolved ARPES data on semiconductors and topological insulators where surface band bending or surface space-charge regions are present, see ref.~\onlinecite{Boschini2024} and references therein. In the presence of a SPV, the measured band structure will appear shifted in energy compared to the equilibrium case - where the shift can be towards higher or lower binding energy depending on the nature of the band bending at the surface (n or p) and the spatial separation of the electron-hole pairs (direction of electric field)~\cite{Rettig2012}. As the SPV is related to a change in the screening of the electric field close to the surface, caused by the photo carriers, it will only affect the energy of the spectrum. Thus, for SPV, no moemtnum ($k$) shift is expected. Furthermore, the time scale for SPV recovery is normally in the order of many picoseconds to hundreds of picoseconds~\cite{Rettig2012,Boschini2024} which does not match the sub-picosecond response observed in our data. 

\begin{figure*}[t]
\includegraphics[width=0.8\textwidth]{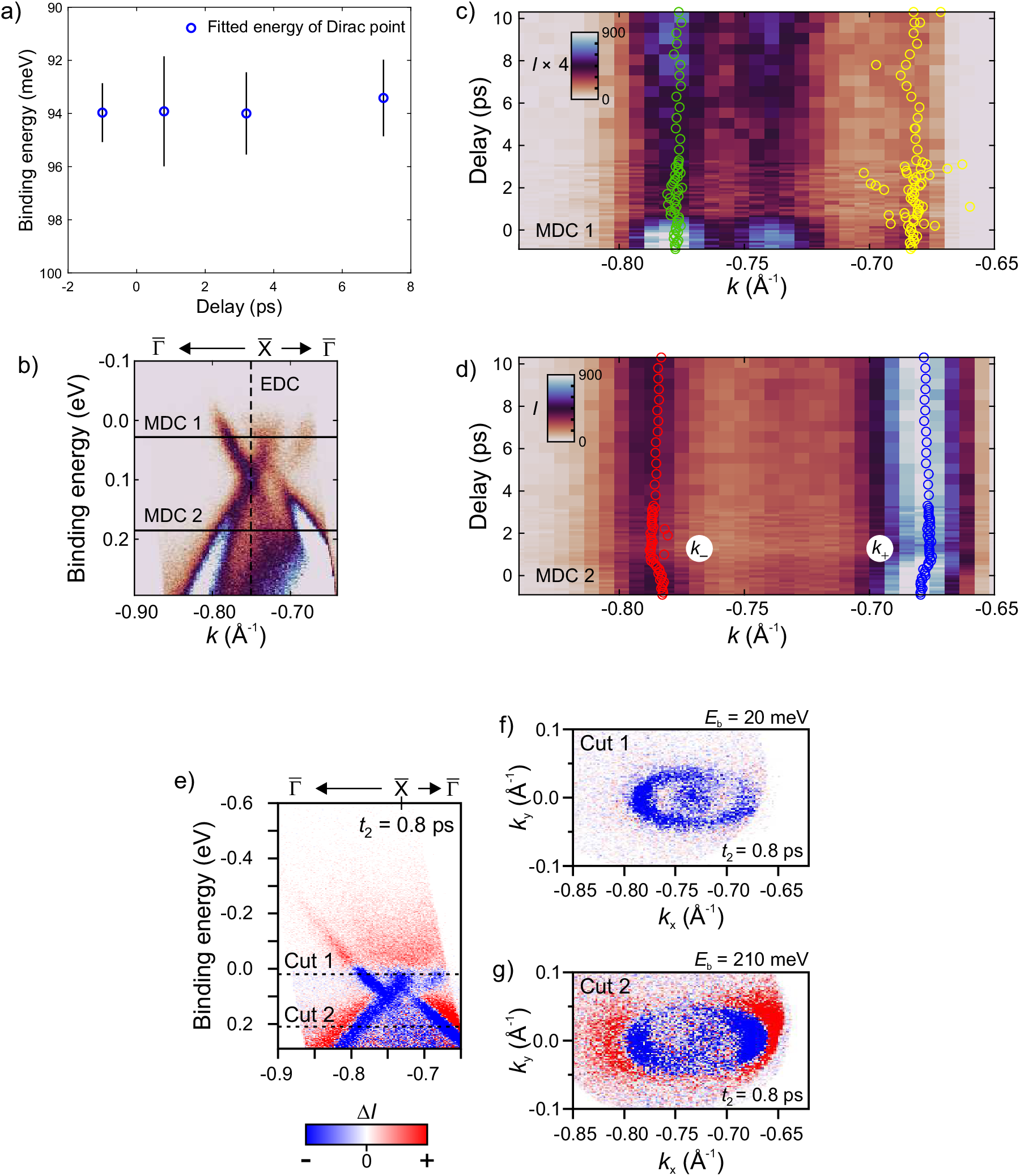}
\caption{\textbf{Stability of Dirac point energy to excitation and behavior of MDC peaks above and below the Dirac energy.} a) Fitted binding energy of the Dirac point deduced from EDC fits. The EDCs are extracted at the momentum position given by the vertical dashed line in panel b) and integrated 0.008~\AA$^{-1}$ in momentum. b) Energy--momentum cut at $t<0$~ps. Horizontal lines labeled MDC 1 and MDC 2 give the binding energy position of the MDCs used for tracing the peak positions of the intersected bands as a function of pump-probe delay time. c)-d) Fitted peak positions of the left and right branches of the topological surface state. e) ARPES difference plot at delay 0.8~ps calculated by using the ARPES spectrum at negative delay as reference. Dashed lines labeled Cut 1 and Cut 2 give the energy positions of the constant energy contours shown in f) and g).}
\label{fig4SI}
\end{figure*}

In Fig.~\ref{fig4SI}a) we follow the binding energy of the Dirac point for different delays deduced from fits of EDCs through the Dirac point. The Dirac point appears to be stable in energy to within a few milli-electronvolts (meV) speaks against the presence of a SPV. Furthermore, the relative shifts $\delta k_{-}$ and $\delta k_{+}$ of the two branches are not symmetric -- indicating a $k$-shift of the entire band structure which is not in line with the SPV effect.

Figure~\ref{fig4SI} c) and d) show fitted peak positions of MDCs extracted at the binding energies indicated in the energy-momentum spectrum in panel b). MDC 2 is the same as shown in the main text in Fig.~4c). If the optical excitation causes a change in the band structure around \Xbar, such that the Dirac point of the surface state moves further away from \Xbar, there should be a shift in the peak positions of the branches intersected by both MDC 1 and MDC 2. Fitted peak positions for both MDCs are shown in panels c) and d), respectively. We notice that the peak shift is clearly observed in MDC 2 whereas for MDC 1 the similar trend is not clearly seen, although the left branch (green markers) shows a small indication of shift in the first two picoseconds. We believe that the absence of a clear peak shift in MDC 1 is caused in part by the challenge to achieve robust peak fits due to the low intensity in this part of the surface state after the excitation. This drop in intensity can be seen in panel c) for positive delays right after $t=0$~ps. Notice the overall lower intensity in MDC 1 compared to MDC 2 in panels c) and d) where the intensity in panel c) has been scaled with a factor of 4.

The significant reduction of intensity in the surface state in the region close to $E_\mathrm{F}$ furthermore causes commonly used intensity difference plots to mask any changes in the energy range between the Dirac point and $E_\mathrm{F}$, as seen in panel e). The constant energy contours corresponding to MDC 1 and MDC 2 are shown in panels f) and g), respectively. This type of difference plots are normally used to visualize changes in photoemission intensity caused by the pump by taking the spectrum at a given delay and subtracting the same spectrum at a negative delay (before excitation). In this particular case, the expected peak shift in MDC 1 is small and thus, due to reduced intensity at positive delays, it will be overshadowed by the significant intensity reduction for this part of the spectrum. See section VI below for further details about the intensity reduction after excitation and on using difference plots to visualize changes in the band structure.

\section{Modeling changes in the band structure due to lattice contraction}

For modeling the effect of a lattice contraction on the band structure of the topological surface state we use a relativistic empirical tight-binding (TB) model~\cite{Lent1986} and solve the eigenvalue problem for a slab geometry consisting of 280 unit cells along the out-of-plane direction. The TB model is parameterized for the sister compound \PbSnTe and the TB parameters for different Sn-concentrations (x) are found by interpolation based on the parameters for PbTe and SnTe through the virtual-crystal approximation (VCA). Qualitatively, the model captures well the experimental band structure, and its Sn-dependence, even for \PbSnSe although being parameterized for a different compound. The different lattice constant for \PbSnTe and \PbSnSe will however position the calculated \Xbar symmetry point of the SBZ at a different $k$-value compared to our experimental data. Additionally, the critical composition for the transition between the NI and TCI phases for \PbSnTe is higher compared to what we see experimentally for \PbSnSe. This means that the compositions used herein to model the band structure do not directly match the experimental values. Nevertheless, the similarities of the two systems allow us to use the TB model of \PbSnTe to extract qualitative information regarding the band structure of \PbSnSe and how it changes with changed Sn content. 

\begin{figure*}
\includegraphics[width=0.9\textwidth]{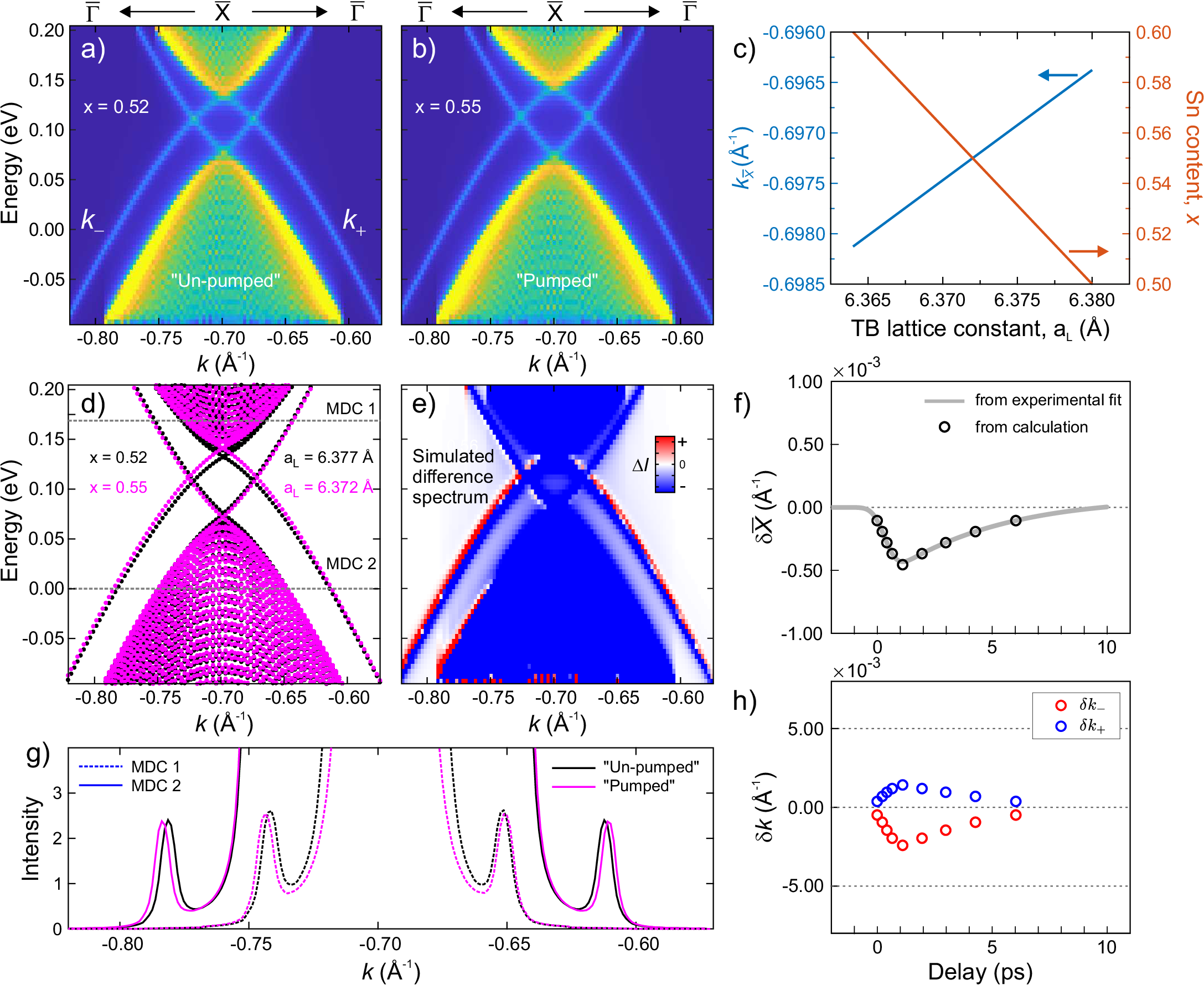}
\caption{\textbf{Modeling changes in the electronic structure after optical excitation using a tight-binding model.} a)-b) Calculated band structures for \PbSnTe with compositions $x=0.52$ and $x=0.55$, respectively. The intensity plots are generated by Lorentzian broadening (along the energy dimension) of the eigenvalues shown in panel d). "Un-pumped" refers to the static band structure at a delay $t<0$~ps while "Pumped" refers to the band structure at the delay where maximum deviation from the static case is observed ($\sim 1.1$~ps). c) The Sn-content ($x$), appearing as the VCA variable for the TB-parameters for varying \PbSnTe compositions, can be used as a proxy for the lattice constant (and hence the momentum position of \Xbar). e) Simulated difference ARPES plot based on the calculated band structures presented in a) and b). f) Relative momentum shift of the \Xbar-point in the SBZ as a function of delay. The gray line is the estimated shift of \Xbar based on the experimental peak fits of MDC 2, as presented in Fig.~4d) in the main text. Circular markers represent corresponding shifts of \Xbar extracted from band structure calculations found by varying $x$ in the TB model. g) MDCs extracted from the simulated intensity plots in a) and b) at the energies indicated by the dashed horizontal lines in d).}
\label{fig6SI}
\end{figure*}

Based on the current XRD data we conclude that the two parameters $T$ (lattice temperature) and $x$ (Sn content) directly influence the lattice constant of \PbSnSe and that changing $T$ or $x$ to a first approximation is equivalent. This enables us to use the $x$-dependence in the TB model as a proxy for the lattice constant ($a_\mathrm{L}$), and therefore also the momentum position of the \Xbar point in the SBZ ($k_{\bar{X}}$), as shown in Fig.~\ref{fig6SI} c). Through this, we incorporate a $a$-dependence in our TB model.

From the experimental data we extract the momentum offset between the Dirac point(s) and \Xbar and compare this with corresponding values given by the TB model for different values of $x$. At $t<0$~ps, the TB model using $x=0.52$ gives the best agreement with the experimental data, corresponding to a lattice constant of $a_\mathrm{L} = 6.377$~\AA in our model. We refer to this as the "un-pumped" case. The calculated eigenvalues are shown in Fig.~\ref{fig6SI} d) as black markers where the Dirac state is clearly visible and distinct from the projected bulk states. Figure~\ref{fig6SI} a) shows the data for $x=0.52$ from d) convoluted along the energy dimension with a Lorentzian kernel ($\mathrm{FWHM} = 10$~meV) in order to mimic energy-momentum intensity plots similar to those observed in ARPES.

From our experimental data, the maximum (relative) shift of the Dirac point away from \Xbar was found to be $4.2\cdot 10^{-3}$\AA$^{-1}$, occurring at a delay of 1.1 ps. In our TB model, by increasing the composition to $x = 0.55$ we can reproduce this shift. In the following we refer to the modeled bandstructre for $x = 0.55$ as the "pumped" case with the eigenvalues and the simulated intensity spectrum shown in panels d) (magenta points) and b), respectively. For this increased value of $x$, the model yields a lattice constant of $a_\mathrm{L} = 6.372$~\AA, indicating a smaller lattice constant for the "pumped" compared to the "un-pumped" lattice.

In panel e) we calculate the intensity difference spectrum based on the two calculated bandstructures in a) and b), taking the "pumped" spectrum minus the "un-pumped" and taking into account the overall intensity variation $I(E)$ across binding energy for the two cases as observed. This means that in the "pumped" case, the intensity of the energy range above the Dirac point experiences a stronger reduction compared to the energy range below the Dirac point (compare Fig.~\ref{fig7SI}, in section VI). The resulting simulated difference ARPES plot in Fig.~\ref{fig6SI}e) qualitatively agrees with its experimental counter part (Fig.~\ref{fig4SI}e)) where an increased intensity is observed outside the lower part of the Dirac cone but not outside the upper part. This masking of changes in the upper part of the Dirac cone is due to the significant intensity reduction in this energy range of the spectrum, rendering the calculated intensity difference overall negative. Computing the simulated difference spectrum without invoking the energy-dependent intensity scaling clearly shows the presence of changes also in the upper part of the spectrum, as demonstrated in Fig.~\ref{fig7SI}d).

As a more direct comparison with the experimental data, we extract MDCs from each of the intensity spectra in a) and b) at energies matching the experimental MDCs presented in Fig.~\ref{fig4SI}. Their energy location are shown by the dashed lines labeled MDC 1 and MDC 2, respectively, in Fig.~\ref{fig6SI}d) and the corresponding intensities in panel g). A close inspection of the MDC peaks in both MDC 1 and MDC 2 in panel g) shows an increased momentum separation of the left and right peaks in the "pumped" case compared to the "un-pumped".

The simulated analogy to the $\delta \bar{X}$ vs. delay plot, presented in Fig.~\ref{fig2SI}, can now be constructed by using the fitted curve of $\delta \bar{X}$ together with the knowledge that the momentum position of \Xbar is determined by the real space lattice constant, $a_\mathrm{L}$. Using the relation between $a_\mathrm{L}$ and $x$ from Fig.~\ref{fig6SI}c) we can then find the corresponding values of $x$ in our model that traces out the $\delta \bar{X}$-curve. The result is shown in panel f). We then extract MDC 2 from the calculated band structures for this series of $x$-values and perform MDC peak fits to get the peak positions for the $k_{-}$ and $k_{+}$ branches respectively. The result is plotted as the relative momentum shift for the two branches as a function of delay in Fig.~\ref{fig6SI}. We notice that the calculations are able to, qualitatively, reproduce the asymmetric momentum shifts $\delta k_{-}$ and $\delta k_{+}$ in agreement with the experimental data presented in Fig.~4d) in the main text (repeated in Fig.~\ref{fig2SI}c) in section III above).

\section{Intensity reduction mask band-structure changes in difference ARPES plots}

Computing difference ARPES spectra -- by subtracting the unpumped spectrum ($t<0$) from the spectrum at a given pump–probe delay ($t>0$) -- is a widely used method for visualizing changes in the electronic structure or population following optical excitation.

However, this approach can obscure subtle spectral changes when the photoexcited population is strongly redistributed. In particular, if the intensity is significantly reduced over a certain energy range, the resulting difference signal in that region becomes strongly negative compared to other energies. This imbalance alters the relative intensity scale across the spectrum, often requiring saturation of the color scale to visualize weaker features. As a consequence, more subtle changes in spectral weight or dispersion can be difficult to resolve in the difference spectrum.

\begin{figure*}
\includegraphics[width=0.6\textwidth]{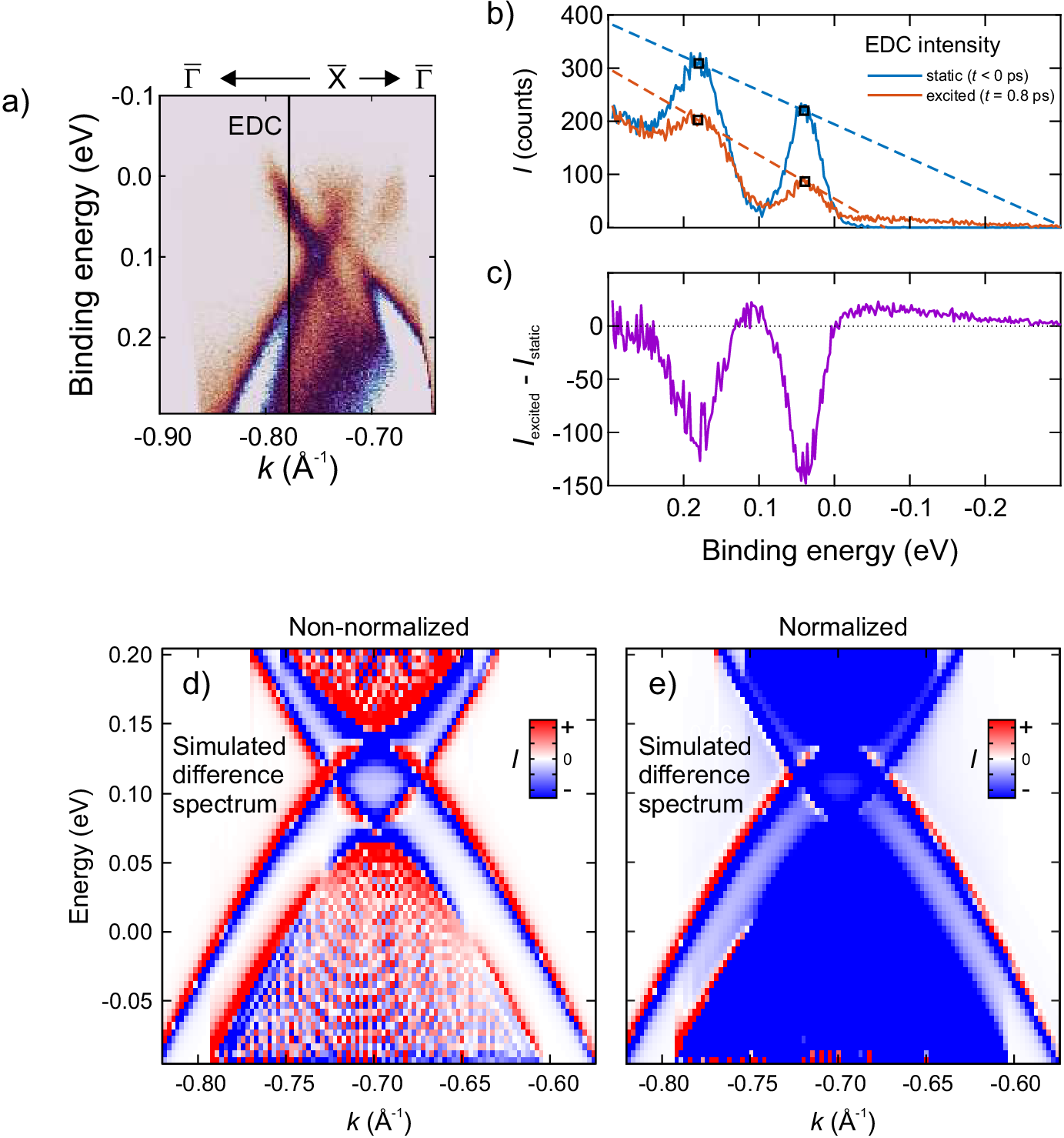}
\caption{\textbf{Demonstration of the need for caution in interpreting difference ARPES spectra} a) Static energy-momentum slice along \GXG with the momentum position of the EDCs shown in b) marked with the vertical line. b)-c) EDCs  and EDC-intensity difference extracted from data for delays $t<0$~ps and $t=0.8$~ps. Dashed lines in b) are linear fits to the peak intensities of the two dominating peaks observed in the EDCs and are used to impose the same $I(E)$ dependence on the simulated data, respectively. d) Simulated difference spectrum for "pumped" with respect to "un-pumped" case without intensity normalization. e) Same difference spectrum as in d), imposing the experimental $I(E)$ dependence from dashed lines in b).}
\label{fig7SI}
\end{figure*}

Figure~\ref{fig7SI} demonstrates this for the present case by simulating such a difference AREPS spectrum based on the calculated band structure. Panel a) shows experimental ARPES data ($t<0$) as a reference, and panel b) presents EDCs ($I(E)$) for $t<0$~ps and $t=0.8$~ps extracted at the momentum position indicated by the vertical line in panel a). Compared to the unpumped case ($t<0$ ps) the EDC at 0.8~ps delay not only shows an overall reduction in intensity, but also a stronger intensity reduction for states approaching the Fermi level. Panel c) displays the calculated intensity difference of the two EDCs in b).

Using the calculated band structures from Fig.~\ref{fig6SI}a) and b), we compute the intensity difference without any intensity normalization and present the results in Fig.~\ref{fig7SI}d). However, normalizing the intensities of the calculated band structures, using the $I(E)$ dependence from Fig.~\ref{fig7SI}b) for the respective delays, we arrive at the difference ARPES plot shown in panel e) of the same figure. From these two simulated difference ARPES spectra we clearly notice that the changes of the band structure occurring in the energy range between the Dirac point(s) and $E_\mathrm{F}$ become less evident when represented in this way. However, the results of panel e) resembles the experimental difference ARPES plot shown in Fig.~\ref{fig4SI}e).

\section{Orbital character of electronic states across the TCI phase transition}

The inversion of the bulk band gap in \PbSnSe can be directly visualized through the inversion of the orbital character of the electronic states in the close vicinity of the band gap. 

\begin{figure*}
\includegraphics[width=0.8\textwidth]{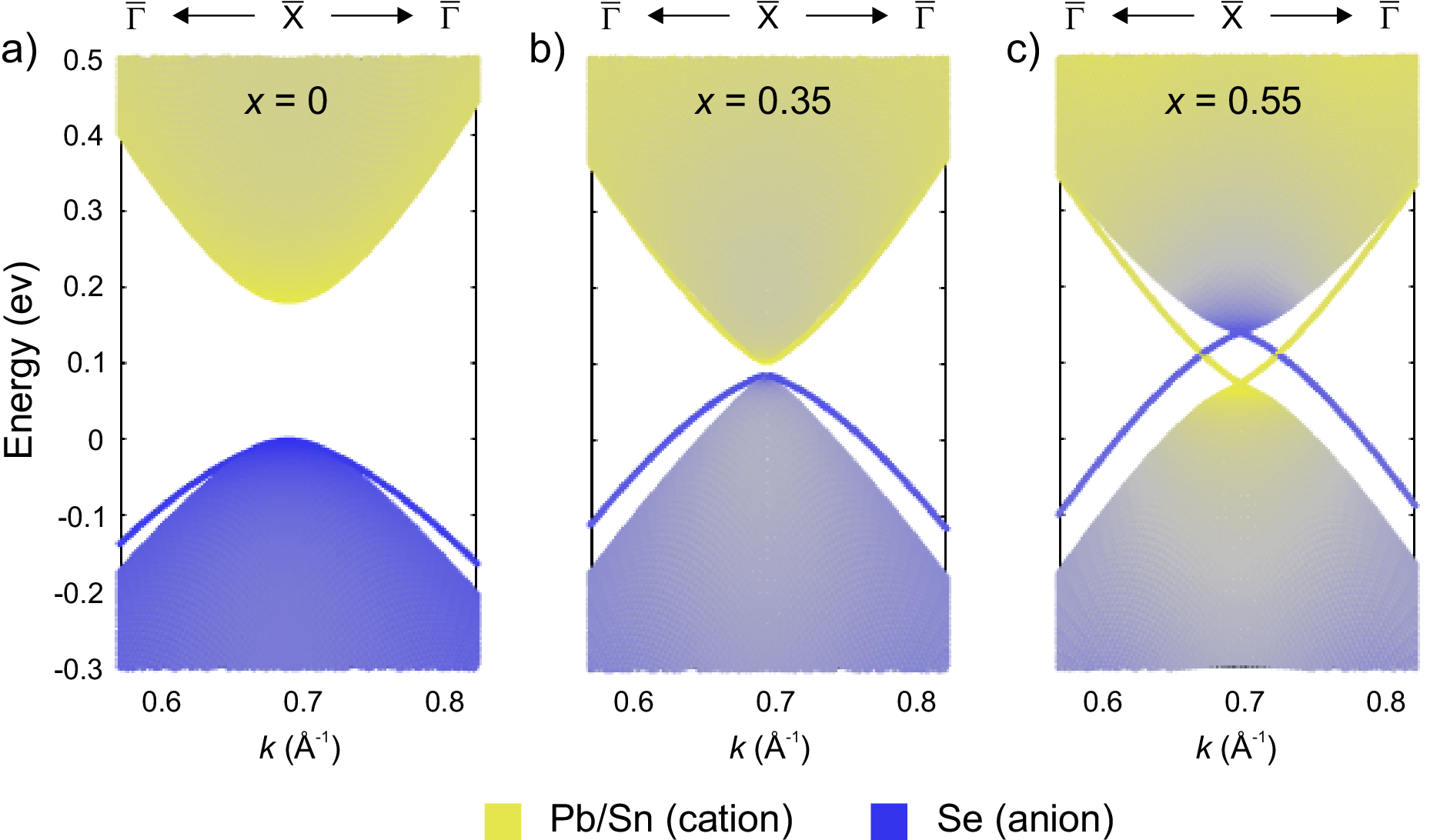}
\caption{\textbf{Orbital character of electronic states in the NI and TCI pahses.} a)-b) For Sn concentrations where \PbSnTe is in the NI phase. c) Orbital character in the TCI phase. Yellow regions indicate states that primarily originate from Pb or Sn orbitals and states with blue color from Se orbitals.}
\label{fig8SI}
\end{figure*}

In \PbSnSe, the band gap is located at the $L$-point in the bulk Brillouin zone, which project to the \Xbar-point in the surface Brillouin zone. From the TB-model, we can extract the orbital character of the electronic states using the calculated eigenvectors. The result is shown in Fig.~\ref{fig8SI} for different Sn-content. For $x=0$, the system is topologically trivial (NI phase) with a positive band gap and the valence and conduction bands have predominantly Se (anionic) and Pb/Sn (cationic) characters, respectively. In the NI phase, no Dirac state is present. Increasing to $x=0.35$, panel b), moves the system closer to the NI-to-TCI phase transition but it still remains in the NI phase. Although significantly reduced, the band gap is still positive and no Dirac state is visible. Increasing the Sn content to $x=0.55$ places the system in the TCI phase, which is manifested by the presence of Dirac states crossing the band gap as well as an inverted orbital character of the states close to the valence and conduction band extrema, as shown in panel c).

\begin{figure*}
\includegraphics[width=0.8\textwidth]{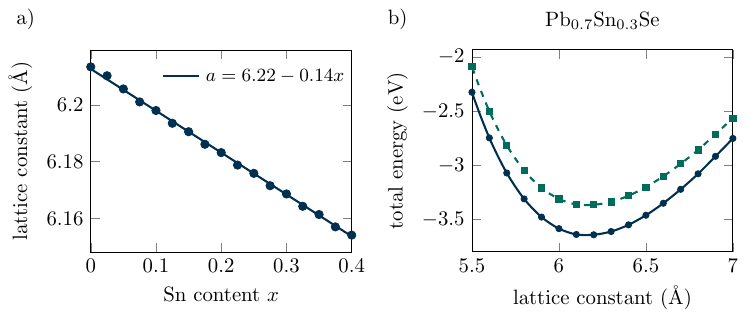}
\caption{\textbf{Structural optimization from ab initio total energy calculations.} a) calculated equilibrium lattice constant for Pb$_{1-x}$Sn$_x$Se crystallizing in the rock salt crystal structure. b) total energy calculations for varying lattice constant and a fixed Sn content $x = 0.3$ for the equilibrium (solid blue, circles) and a non-equilibrium (dashed green, squares) electron distribution. The non-equilibrium distribution corresponds to an excitation with 0.3 electrons per primitive unit cell.}
\label{fig9SI}
\end{figure*}

\section{ab initio crystal lattice optimization}

Figure~\ref{fig9SI}a) presents the optimized equilibrium lattice constant across a relevant range of Sn concentrations determined by DFT. The linear decrease of $a$ with increased $x$ is observed in agreement with experiments. Panel b) in the same figure shows the principle of total energy minimization with respect to $a$ for different values of $\delta$. See the Methods section in the main text for details about the calculations.

\bibliography{references_SI.bib}